%% file: main.tex
\documentclass[journal]{IEEEtran}

\RequirePackage{xcolor}
\PassOptionsToPackage{detect-all, per-mode=symbol, range-units=single, range-phrase=~to~}{siunitx}
\RequirePackage{siunitx}
\PassOptionsToPackage{acronyms,nohypertypes={acronym},nomain}{glossaries}
\RequirePackage{glossaries}
\RequirePackage[pscoord]{eso-pic}
\RequirePackage{multirow}
\RequirePackage{makecell}
\RequirePackage{threeparttable}
\RequirePackage{booktabs}
\RequirePackage{colortbl}
\RequirePackage[verbose]{placeins}
\RequirePackage{fancyhdr}
\RequirePackage[hidelinks]{hyperref}
\RequirePackage{MnSymbol}
\RequirePackage{wasysym}
\RequirePackage{xspace}
\RequirePackage{lipsum}
\RequirePackage{soul}
\RequirePackage{calc}
\RequirePackage{float}
\RequirePackage{threeparttable}
\RequirePackage{booktabs}
\RequirePackage{colortbl}
\RequirePackage{tabularx}
\RequirePackage{makecell}
\RequirePackage{pifont}
\RequirePackage{arydshln}

\RequirePackage{amsmath,amsfonts}
\RequirePackage{algorithmic}
\RequirePackage{algorithm}
\RequirePackage{array}
\RequirePackage{subcaption}
\RequirePackage{url}
\RequirePackage{verbatim}
\RequirePackage{graphicx}

\RequirePackage{orcidlink}
\RequirePackage[noabbrev,capitalise]{cleveref}

\RequirePackage{cite}

\newcommand{\glsu}[1]{\glsunset{#1}\gls{#1}}

\newcommand{\etal}{\emph{et al.}}
\newcommand{\x}{$\times$}

\DeclareSIUnit{\x}{\!\ensuremath{\times}}
\DeclareSIUnit\bit{b}
\DeclareSIUnit\GE{GE}
\DeclareSIUnit\kGE{\kilo\GE}
\DeclareSIUnit\MGE{\mega\GE}
\newlength\myheight
\newlength\mydepth
\settototalheight\myheight{Xygp}
\newcommand*\circnum[1]{\tikz[baseline=(char.base)]{%
            \node[white,shape=circle,fill=ieee-dark-black-100,draw,inner sep=1pt] (char) {\color{ieee-bright-white-100}\sffamily #1};}}

\definecolor{ieee-bright-dblue-100}{rgb}{0.0, 0.3828, 0.6055}
\definecolor{ieee-bright-dblue-80}{rgb}{0.0, 0.4883, 0.6797}
\definecolor{ieee-bright-dblue-60}{rgb}{0.3633, 0.6094, 0.7617}
\definecolor{ieee-bright-dblue-40}{rgb}{0.5898, 0.7383, 0.8398}
\definecolor{ieee-bright-dblue-20}{rgb}{0.8906, 0.8984, 0.9219}
\definecolor{ieee-bright-red-100}{rgb}{0.7266, 0.0469, 0.1836}
\definecolor{ieee-bright-red-80}{rgb}{0.832, 0.3164, 0.3281}
\definecolor{ieee-bright-red-60}{rgb}{0.8906, 0.4922, 0.4805}
\definecolor{ieee-bright-red-40}{rgb}{0.9336, 0.6562, 0.6406}
\definecolor{ieee-bright-red-20}{rgb}{0.9688, 0.8203, 0.8125}
\definecolor{ieee-bright-orange-100}{rgb}{0.9961, 0.6367, 0.0}
\definecolor{ieee-bright-orange-80}{rgb}{0.9844, 0.6953, 0.3125}
\definecolor{ieee-bright-orange-60}{rgb}{0.9883, 0.7695, 0.4844}
\definecolor{ieee-bright-orange-40}{rgb}{0.9922, 0.8359, 0.6562}
\definecolor{ieee-bright-orange-20}{rgb}{0.9961, 0.9219, 0.8164}
\definecolor{ieee-bright-yellow-100}{rgb}{0.9961, 0.8164, 0.0}
\definecolor{ieee-bright-yellow-80}{rgb}{0.9961, 0.8477, 0.2148}
\definecolor{ieee-bright-yellow-60}{rgb}{0.9961, 0.875, 0.4492}
\definecolor{ieee-bright-yellow-40}{rgb}{0.9961, 0.9062, 0.6328}
\definecolor{ieee-bright-yellow-20}{rgb}{0.9961, 0.9531, 0.8125}
\definecolor{ieee-bright-lgreen-100}{rgb}{0.4688, 0.7422, 0.125}
\definecolor{ieee-bright-lgreen-80}{rgb}{0.5742, 0.7852, 0.332}
\definecolor{ieee-bright-lgreen-60}{rgb}{0.6875, 0.8398, 0.5039}
\definecolor{ieee-bright-lgreen-40}{rgb}{0.793, 0.8906, 0.6641}
\definecolor{ieee-bright-lgreen-20}{rgb}{0.8945, 0.9414, 0.8281}
\definecolor{ieee-bright-dgreen-100}{rgb}{0.0, 0.5156, 0.2383}
\definecolor{ieee-bright-dgreen-80}{rgb}{0.1641, 0.6055, 0.3867}
\definecolor{ieee-bright-dgreen-60}{rgb}{0.3906, 0.6953, 0.5234}
\definecolor{ieee-bright-dgreen-40}{rgb}{0.6094, 0.8008, 0.6719}
\definecolor{ieee-bright-dgreen-20}{rgb}{0.8047, 0.8945, 0.8359}
\definecolor{ieee-bright-purple-100}{rgb}{0.5938, 0.1133, 0.5898}
\definecolor{ieee-bright-purple-80}{rgb}{0.6992, 0.3281, 0.668}
\definecolor{ieee-bright-purple-60}{rgb}{0.7812, 0.4961, 0.7461}
\definecolor{ieee-bright-purple-40}{rgb}{0.8555, 0.6602, 0.8281}
\definecolor{ieee-bright-purple-20}{rgb}{0.9219, 0.8281, 0.9023}
\definecolor{ieee-bright-lblue-100}{rgb}{0.0, 0.6094, 0.6484}
\definecolor{ieee-bright-lblue-80}{rgb}{0.0, 0.6797, 0.7188}
\definecolor{ieee-bright-lblue-60}{rgb}{0.2109, 0.75, 0.7812}
\definecolor{ieee-bright-lblue-40}{rgb}{0.5469, 0.8242, 0.8438}
\definecolor{ieee-bright-lblue-20}{rgb}{0.7695, 0.918, 0.9219}
\definecolor{ieee-bright-cyan-100}{rgb}{0.0, 0.707, 0.8828}
\definecolor{ieee-bright-cyan-80}{rgb}{0.0, 0.7227, 0.9453}
\definecolor{ieee-bright-cyan-60}{rgb}{0.2656, 0.7812, 0.957}
\definecolor{ieee-bright-cyan-40}{rgb}{0.5547, 0.8438, 0.9688}
\definecolor{ieee-bright-cyan-20}{rgb}{0.7773, 0.9141, 0.9805}
\definecolor{ieee-bright-white-100}{rgb}{0.9961, 0.9961, 0.9961}
\definecolor{ieee-bright-white-80}{rgb}{0.9961, 0.9961, 0.9961}
\definecolor{ieee-bright-white-60}{rgb}{0.9961, 0.9961, 0.9961}
\definecolor{ieee-bright-white-40}{rgb}{0.9961, 0.9961, 0.9961}
\definecolor{ieee-bright-white-20}{rgb}{0.9961, 0.9961, 0.9961}
\definecolor{ieee-dark-red-100}{rgb}{0.5234, 0.1211, 0.2539}
\definecolor{ieee-dark-red-80}{rgb}{0.6445, 0.2812, 0.3828}
\definecolor{ieee-dark-red-60}{rgb}{0.7422, 0.4727, 0.5234}
\definecolor{ieee-dark-red-40}{rgb}{0.832, 0.6445, 0.6758}
\definecolor{ieee-dark-red-20}{rgb}{0.918, 0.8203, 0.832}
\definecolor{ieee-dark-orange-100}{rgb}{0.9062, 0.4648, 0.1328}
\definecolor{ieee-dark-orange-80}{rgb}{0.9648, 0.5664, 0.3164}
\definecolor{ieee-dark-orange-60}{rgb}{0.9766, 0.6758, 0.4805}
\definecolor{ieee-dark-orange-40}{rgb}{0.9844, 0.7773, 0.6523}
\definecolor{ieee-dark-orange-20}{rgb}{0.9922, 0.8789, 0.8125}
\definecolor{ieee-dark-yellow-100}{rgb}{0.9961, 0.7773, 0.1719}
\definecolor{ieee-dark-yellow-80}{rgb}{0.9961, 0.8086, 0.375}
\definecolor{ieee-dark-yellow-60}{rgb}{0.9961, 0.875, 0.4492}
\definecolor{ieee-dark-yellow-40}{rgb}{0.9961, 0.8984, 0.6875}
\definecolor{ieee-dark-yellow-20}{rgb}{0.9961, 0.9453, 0.8438}
\definecolor{ieee-dark-lgreen-100}{rgb}{0.3945, 0.5508, 0.0938}
\definecolor{ieee-dark-lgreen-80}{rgb}{0.5078, 0.6289, 0.293}
\definecolor{ieee-dark-lgreen-60}{rgb}{0.6367, 0.7188, 0.4688}
\definecolor{ieee-dark-lgreen-40}{rgb}{0.7539, 0.8047, 0.6367}
\definecolor{ieee-dark-lgreen-20}{rgb}{0.875, 0.9023, 0.8125}
\definecolor{ieee-dark-dgreen-100}{rgb}{0.0, 0.3867, 0.2539}
\definecolor{ieee-dark-dgreen-80}{rgb}{0.1836, 0.5, 0.3906}
\definecolor{ieee-dark-dgreen-60}{rgb}{0.3984, 0.6172, 0.5273}
\definecolor{ieee-dark-dgreen-40}{rgb}{0.5938, 0.7422, 0.6758}
\definecolor{ieee-dark-dgreen-20}{rgb}{0.793, 0.8711, 0.8359}
\definecolor{ieee-dark-purple-100}{rgb}{0.4648, 0.1445, 0.5117}
\definecolor{ieee-dark-purple-80}{rgb}{0.5898, 0.3242, 0.6016}
\definecolor{ieee-dark-purple-60}{rgb}{0.6914, 0.4883, 0.6953}
\definecolor{ieee-dark-purple-40}{rgb}{0.7969, 0.6523, 0.793}
\definecolor{ieee-dark-purple-20}{rgb}{0.8945, 0.8203, 0.8945}
\definecolor{ieee-dark-cyan-100}{rgb}{0.0, 0.4492, 0.4648}
\definecolor{ieee-dark-cyan-80}{rgb}{0.0, 0.5469, 0.5664}
\definecolor{ieee-dark-cyan-60}{rgb}{0.3047, 0.6602, 0.668}
\definecolor{ieee-dark-cyan-40}{rgb}{0.5586, 0.7695, 0.7734}
\definecolor{ieee-dark-cyan-20}{rgb}{0.7734, 0.8789, 0.8789}
\definecolor{ieee-dark-dblue-100}{rgb}{0.0, 0.1562, 0.332}
\definecolor{ieee-dark-dblue-80}{rgb}{0.1797, 0.3008, 0.4609}
\definecolor{ieee-dark-dblue-60}{rgb}{0.3828, 0.4609, 0.5859}
\definecolor{ieee-dark-dblue-40}{rgb}{0.5781, 0.6289, 0.7188}
\definecolor{ieee-dark-dblue-20}{rgb}{0.7852, 0.8047, 0.8555}
\definecolor{ieee-dark-grey-100}{rgb}{0.457, 0.4688, 0.4805}
\definecolor{ieee-dark-grey-80}{rgb}{0.5625, 0.5625, 0.5742}
\definecolor{ieee-dark-grey-60}{rgb}{0.6641, 0.6641, 0.6758}
\definecolor{ieee-dark-grey-40}{rgb}{0.7734, 0.7695, 0.7773}
\definecolor{ieee-dark-grey-20}{rgb}{0.8789, 0.8828, 0.8828}
\definecolor{ieee-dark-black-100}{rgb}{0.0, 0.0, 0.0}
\definecolor{ieee-dark-black-80}{rgb}{0.3438, 0.3477, 0.3555}
\definecolor{ieee-dark-black-60}{rgb}{0.5, 0.5078, 0.5195}
\definecolor{ieee-dark-black-40}{rgb}{0.6523, 0.6602, 0.6719}
\definecolor{ieee-dark-black-20}{rgb}{0.8164, 0.8242, 0.8281}

\IfFileExists{identifier.tex}{
    \def\reviewpass{SubmittedArxivAccepted-bed55018d25905fd}

}{
    \def\reviewpass{Overleaf Version - Do not distribute!}
}

\def\thetitle{{\axirealm}: Safe, Modular and Lightweight Traffic Monitoring and Regulation \lb{for} Heterogeneous Mixed-Criticality Systems}
\def\thetitlesingleline{{\axirealm}: Safe, Modular and Lightweight Traffic Monitoring and Regulation \lb{for} Heterogeneous Mixed-Criticality Systems}

\def\showdisclaimer{}

\ifx\showrebuttal\undefined
    
\else
    
\fi

\newacronym{sota}{SOTA}{state-of-the-art}
\newacronym{fpga}{FPGA}{field programmable gate array}
\newacronym{asic}{ASIC}{application-specific integrated circuit}
\newacronym{fub}{FUB}{functional unit block}
\newacronym{vv}{V\&V}{validation and verification}
\newacronym{gpp}{GPP}{general purpose processor}
\newacronym{tid}{tID}{transaction ID}
\newacronym{cots}{COTS}{commercial-off-the-shelf}

\newacronym{hpc}{HPC}{high performance computing}
\newacronym{ml}{ML}{machine learning}
\newacronym{isa}{ISA}{instruction set architecture}
\newacronym{fp}{FP}{floating-point}
\newacronym{dl}{DL}{deep learning}
\newacronym{la}{LA}{linear algebra}
\newacronym{ip}{IP}{intellectual property}
\newacronym[firstplural=systems-on-chip (SoCs)]{soc}{SoC}{system-on-chip}
\newacronym{mpsoc}{MPSoC}{multi-processor system-on-chip}
\newacronym[firstplural=networks-on-chip (NoCs)]{noc}{NoC}{network-on-chip}
\newacronym{hw}{HW}{hardware}
\newacronym{sw}{SW}{software}
\newacronym{swapc}{SWaP-C}{space, weight, power, and cost}
\newacronym{mcp}{MCP}{multi-core processor}
\newacronym{rr}{RR}{round-robin}
\newacronym{tdma}{TDMA}{time division multiple access}
\newacronym{aces}{ACES}{Autonomous driving, Connectivity, Electrification, and Shared mobility} %
\newacronym{sdv}{SDV}{software-defined vehicles}

\newacronym{mac}{MAC}{multiply-accumulate}
\newacronym{fem}{FEM}{finite element analysis}
\newacronym{simd}{SIMD}{single-instruction, multiple-data}
\newacronym{rtl}{RTL}{register transfer level}
\newacronym{dlt}{DLT}{data layout transform}

\newacronym{fifo}{FIFO}{first in, first out}
\newacronym{fu}{FU}{functional unit}
\newacronym{alu}{ALU}{arithmetic logic unit}
\newacronym{fpu}{FPU}{floating-point unit}
\newacronym{ssr}{SSR}{stream semantic register}
\newacronym{issr}{ISSR}{indirection stream semantic register}
\newacronym{tcdm}{TCDM}{tightly-coupled data memory}
\newacronym{dma}{DMA}{direct memory access}
\newacronym{sm}{SM}{streaming multiprocessor}
\newacronym{vlsu}{VLSU}{vector load-store unit}
\newacronym{dsa}{DSA}{domain-specific accelerator}
\newacronym{ha}{HA}{hardware accelerator}
\newacronym{fsm}{FSM}{finite state machine}
\newacronym{llc}{LLC}{last-level cache}
\newacronym{d2d}{D2D}{die-to-die}
\newacronym{dram}{DRAM}{dynamic random access memory}
\newacronym{spm}{SPM}{scratchpad memory}
\newacronym{rf}{RF}{register file}
\newacronym{mmu}{MMU}{memory management unit}
\newacronym{pmp}{MMU}{physical memory protection unit}
\newacronym{l2}{L2}{level-two}
\newacronym{clic}{CLIC}{core-local interrupt controller}
\newacronym{eth}{ETH}{Ethernet}

\newacronym{os}{OS}{operating system}

\newacronym{spvv}{SpVV}{sparse vector-vector multiplication}
\newacronym{spmv}{SpMV}{sparse vector-matrix multiplication}
\newacronym{spmm}{SpMM}{sparse matrix-matrix multiplication}
\newacronym{csrmv}{CsrMV}{CSR matrix-vector multiplication}
\newacronym{csrmm}{CsrMM}{CSR matrix-matrix multiplication}

\newacronym{csf}{CSF}{compressed sparse fiber}
\newacronym{csr}{CSR}{compressed sparse rows}
\newacronym{csc}{CSC}{compressed sparse columns}
\newacronym{bcsr}{BCSR}{blocked compressed sparse rows}

\newacronym{axi4}{AXI4}{Advanced eXtensible Interface 4}
\newacronym{amba}{AMBA}{Advanced Microcontroller Bus Architecture}
\newacronym{sram}{SRAM}{static random-access memory}

\newacronym{wcet}{WCET}{worst-case execution time}
\newacronym{wcdt}{WCDT}{worst-case detection time}
\newacronym{rtunit}{REALM unit}{real-time regulation and traffic monitoring unit}
\newacronym{mtunit}{M\&R unit}{monitoring and regulation unit}
\newacronym{cps}{CPS}{cyber-physical system}
\newacronym{crtes}{CRTES}{critical real-time embedded system}
\newacronym{heicps}{He-iCPS}{heterogeneous integrated cyber-physical system}
\newacronym{ecu}{ECU}{electronic control unit}
\newacronym{mcs}{MCS}{mixed criticality system}
\newacronym{ima}{IMA}{integrated modular avionics}
\newacronym{adas}{ADAS}{Advanced Driver Assistance System} %
\newacronym{axirealm}{AXI-REALM}{AXI real-time regulation and traffic monitoring}
\newacronym{mpam}{MPAM}{memory system resource partitioning and monitoring}
\newacronym{dos}{DoS}{denial of service}
\newacronym{hwrot}{HWRoT}{hardware root of trust}
\newacronym{pcs}{PCS}{power controller subsystem}
\newacronym{sil}{SIL}{safety integrity level}
\newacronym{tmu}{TMU}{transaction monitor unit}

\newacronym{ps}{PS}{per-system}
\newacronym{pu}{PU}{per-unit}
\newacronym{pur}{PUR}{per unit and region}

\newcommand{\gfs}{{GlobalFoundries'}}
\newcommand{\gftech}{{GF12LP+}}
\newcommand{\dc}{{Synopsys} {Design} {Compiler} {NXT} 2023.12}
\newcommand{\pt}{{Synopsys} {PrimeTime} 2022.03}
\newcommand{\riscv}{\mbox{RISC-V}}

\ifx\showacrcols\undefined
    \newcommand{\carfield}{{Carfield}}
\else
    \newcommand{\carfield}{\textcolor{ieee-bright-cyan-100}{Hermes}}
\fi

\ifx\showacrcols\undefined
    \newcommand{\axirealm}{{AXI-REALM}}
    \newcommand{\irealm}{i{realm}}
    \newcommand{\erealm}{e{realm}}
\else
    \newcommand{\axirealm}{\textcolor{ieee-bright-orange-100}{{AXI-REALM}}}
    \newcommand{\irealm}{\textcolor{ieee-bright-lgreen-100}{i{realm}}}
    \newcommand{\erealm}{\textcolor{ieee-bright-purple-100}{e{realm}}}
\fi

\newcommand{\cmark}{\ding{51}}%
\newcommand{\xmark}{\ding{55}}%

\begin{document}

\ifx\showdisclaimer\undefined
\else
\AddToShipoutPictureBG*{%
  \AtPageUpperLeft{%
    \hspace{\paperwidth}%
    \raisebox{-\baselineskip}{%
      \makebox[-35pt][r]{\footnotesize{
        \copyright~2025~IEEE. Personal use of this material is permitted. %
        Permission from IEEE must be obtained for all other uses, in any current or future media, including
      }}
}}}%

\AddToShipoutPictureBG*{%
  \AtPageUpperLeft{%
    \hspace{\paperwidth}%
    \raisebox{-2\baselineskip}{%
      \makebox[-37pt][r]{\footnotesize{
        reprinting/republishing this material for advertising or promotional purposes, creating new collective works, for resale or redistribution to servers or lists, or
      }}
}}}%

\AddToShipoutPictureBG*{%
  \AtPageUpperLeft{%
    \hspace{\paperwidth}%
    \raisebox{-3\baselineskip}{%
      \makebox[-185pt][r]{\footnotesize{
       reuse of any copyrighted component of this work in other works.
      }}
}}}%
\fi

\title{\thetitle}
\ifx\showrevision\undefined
    \newcommand{\todo}[1]{{#1}}
\else
    \newcommand{\todo}[1]{{\textcolor{red}{#1}}}
    \AddToShipoutPictureFG{%
        \put(%
            8mm,%
            \paperheight-1.5cm%
            ){\vtop{{\null}\makebox[0pt][c]{%
                \rotatebox[origin=c]{90}{%
                    \huge\textcolor{red!75}{\reviewpass}%
                }%
            }}%
        }%
    }
    \AddToShipoutPictureFG{%
        \put(%
            \paperwidth-6mm,%
            \paperheight-1.5cm%
            ){\vtop{{\null}\makebox[0pt][c]{%
                \rotatebox[origin=c]{90}{%
                    \huge\textcolor{red!30}{ETH Zurich - Unpublished - Confidential - Draft - Copyright Thomas, Ale 2025}%
                }%
            }}%
        }%
    }
\fi

\ifx\showfeedback\undefined
    \newcommand{\lb}[1]{#1}
    \newcommand{\lbb}[1]{#1}
    \newcommand{\lbc}[1]{#1}
    \newcommand{\ab}[1]{#1}
    \newcommand{\fr}[1]{#1}
    \newcommand{\dr}[1]{#1}
\else
    \newcommand{\lb}[1]{{\textcolor{ieee-bright-lblue-100}{#1}}}
    \newcommand{\lbb}[1]{{\textcolor{ieee-bright-lblue-80}{#1}}}
    \newcommand{\lbc}[1]{{\textcolor{ieee-bright-lblue-60}{#1}}}
    \newcommand{\ab}[1]{{\textcolor{ieee-bright-lgreen-100}{#1}}}
    \newcommand{\fr}[1]{{\textcolor{ieee-bright-orange-100}{#1}}}
    \newcommand{\dr}[1]{{\textcolor{ieee-bright-purple-100}{#1}}}
\fi

\ifx\showrebuttal\undefined
    \newcommand{\rev}[1]{#1}
\else
    \newcommand{\rev}[1]{{\textcolor{ieee-bright-lblue-100}{#1}}}
\fi

\ifx\showrebuttal\undefined
    \newcommand{\revdel}[1]{}
\else
    \newcommand{\revdel}[1]{\textcolor{ieee-bright-red-100}{\st{#1}}}
\fi

\ifx\showrebuttal\undefined
    \newcommand{\revrep}[2]{#2}
\else
    \newcommand{\revrep}[2]{\revdel{#1} \rev{#2}}
\fi

\ifx\showrebuttal\undefined
    \newcommand{\revprg}[1]{}
\else
    \newcommand{\revprg}[1]{\hspace{-0.5ex}\textcolor{ieee-bright-red-100}{\scalebox{.2}[1.5]{$\blacksquare$}}\hspace{-0.5ex}}
\fi

\author{
    Thomas~Benz\orcidlink{0000-0002-0326-9676}\IEEEauthorrefmark{10},~\IEEEmembership{Graduate Student Member, IEEE},
    Alessandro~Ottaviano\orcidlink{0009-0000-9924-3536}\IEEEauthorrefmark{10},~\IEEEmembership{Graduate Student Member, IEEE},
    Chaoqun~Liang\orcidlink{0009-0008-0556-7758},~\IEEEmembership{Graduate Student Member, IEEE},
    Robert~Balas\orcidlink{0000-0002-7231-9315},~\IEEEmembership{Graduate Student Member, IEEE},
    Angelo~Garofalo\orcidlink{0000-0002-7495-6895},~\IEEEmembership{Member, IEEE},
    Francesco~Restuccia\orcidlink{0000-0001-6955-1888},~\IEEEmembership{Member, IEEE},
    Alessandro~Biondi\orcidlink{0000-0002-6625-9336},~\IEEEmembership{Member, IEEE},
    Davide~Rossi\orcidlink{0000-0002-0651-5393},~\IEEEmembership{Senior Member, IEEE},
    and~Luca~Benini\orcidlink{0000-0001-8068-3806},~\IEEEmembership{Fellow,~IEEE}%
    \vspace{-0.5cm}
    \IEEEcompsocitemizethanks{%
    \IEEEauthorrefmark{10} Both authors contributed equally to this research.\protect\\
    \IEEEcompsocthanksitem T.~Benz, A.~Ottaviano, R.~Balas, A.~Garofalo and L.~Benini are with the Integrated Systems Laboratory (IIS), ETH Zurich, Switzerland.\protect\\
    E-mail: \{tbenz,aottaviano\}@ethz.ch\protect\\
    Francesco~Restuccia is with the Department of Computer Science and Engineering, UC San Diego, San Diego, CA USA.\protect\\
    Alessandro~Biondi is with the Department of Excellence in Robotics \& AI, Scuola Superiore Sant'Anna, Pisa, Italy.\protect\\
    C.~Liang, A.~Garofalo, D.~Rossi, and L.~Benini are with the Department of Electrical, Electronic, and Information Engineering (DEI), University of Bologna, Bologna, Italy.\protect\\
    }%
}

\ifx\showdisclaimer\undefined
\markboth{IEEE Transactions on Computers}%
{Benz \MakeLowercase{\etal}: \thetitlesingleline}
\else
\fi

\maketitle

\begin{abstract}
The automotive industry is transitioning from federated, homogeneous, interconnected devices to integrated, heterogeneous, mixed-criticality systems (MCS).
This leads to challenges in \ab{achieving} timing predictability techniques due to access contention on shared resources, which can be mitigated using hardware-based spatial and temporal isolation techniques.
Focusing on the interconnect as \fr{the point of access for shared resources}, we propose {\axirealm}, a lightweight, modular, technology-independent, and open-source real-time extension to AXI4 interconnects.
{\axirealm} uses a budget-based mechanism \ab{enforced} on periodic time windows and transfer fragmentation to provide fair arbitration, \lbb{coupled with} execution predictability on real-time workloads.
{\axirealm} features a comprehensive bandwidth and latency monitor at both the ingress and egress of the interconnect system.
Latency information is \lbb{also} used to detect and reset malfunctioning subordinates, preventing missed deadlines.
We provide a detailed cost assessment in a \SI{12}{\nano\metre} node and an end-to-end case study implementing {\axirealm} into an open-source MCS, incurring an area overhead of less than \SI{2}{\percent}. %
When running \lbb{a mixed-criticality workload, with a time-critical application sharing the interconnect with non-critical applications}, we demonstrate that the critical application can achieve up to \SI{68.2}{\percent} of the isolated performance by enforcing fairness on the interconnect traffic through burst fragmentation, thus \lb{reducing the subordinate access latency by} \lbb{up to 24 \lbc{times}}.
Near-ideal performance\lb{,} (above \SI{95}{\%} \lb{of the isolated performance}) can be achieved by distributing the available bandwidth in favor of the critical application.

\end{abstract}


\glsresetall

\section{Introduction}
\label{sec:intro}

The current trend in industrial domains such as automotive, robotics, and aerospace is towards autonomy, connectivity, and electrification, significantly increasing the demand for onboard computing power and communication infrastructure, thus driving a paradigm shift in their design~\cite{src2023mapt, burkacky2023getting, mutschler2024automotive, jiang2023towardshardreal, kasarapu2025performanceande, MCKINSEY_AUTOMOTIVE_SURVEY}.

A clear example is the automotive domain, where the traditional approach --- relying on hundreds of embedded real-time \glspl{ecu} distributed throughout the vehicle --- \lbb{cannot} meet the growing compute demands and complicates cable harness management, impacting \gls{swapc}~\cite{MCKINSEY_AUTOMOTIVE_SURVEY, lim2024aframeworkforde, pinto2019virtualizationo}.
This paradigm cannot support the rapid shift toward \lb{\gls{aces}}, which is laying the foundation of \lb{\gls{adas}} and \glspl{sdv}~\cite{burkacky2023getting}.
Hence, integrated, interconnected \emph{zonal} and \emph{domain} architectures are becoming the preferred replacements for discrete \glspl{ecu}, as they deliver the flexibility and compute \lbc{capability} required for \gls{aces} mobility and the \gls{swapc} problem~\cite{jang2023designofahybrid, pinto2019virtualizationo}.

These architectures are heterogeneous \glspl{mcs}~\cite{MCS_REVIEW_2}.
They comprise general-purpose and \lbc{domain-specific sub-systems} with diverse real-time and \lbb{specialized} computing requirements that execute concurrently on the same \lbc{silicon die}, sharing communication, storage\ab{, and micro-architectural} resources~\cite{majumder2020partaaarealtime, sá2022afirstlookatris}.
Some \lbc{subsystems} handle hard safety- and time-critical workloads, such as engine, brake, and cruise control~\cite{MCKINSEY_AUTOMOTIVE_SURVEY, RETIS_BANDWIDTH_RESERVATION, AXIICRT_ARM}, while others run less time-critical but computationally demanding tasks like perception pipelines, infotainment, and commodity applications~\cite{MCKINSEY_AUTOMOTIVE_SURVEY}.

Time- and safety-critical \lbc{tasks} require \fr{strict real-time guarantees}, ensured through \ab{time-predictable run-time mechanisms, composable timing analysis}, and safety assessments~\cite{SOA_WCET}.
However, in heterogeneous \glspl{mcs}, this process \lb{is complicated by} the increased interference generated by multiple domains contending for shared hardware resources on the same platform~\cite{RETIS_BANDWIDTH_RESERVATION}.
This additional contention may introduce unpredictable behavior during the system's execution, causing possible deadline misses for time- and safety-critical tasks~\cite{RETIS_AXI_HYPER_CONNECT, CAST32A_POSITION_PAPER}.
To preserve the timing behavior of the system under known and predictable bounds, techniques such as spatial and temporal isolation become a prerequisite, as they enhance the \emph{observability} and \emph{controllability} of shared \gls{hw} resources~\cite{AUTOMOTIVE_PREDICTABLE, jiang2023towardshardreal}.
\fr{The \emph{interconnect} in modern \glspl{soc} is of particular \lbc{concern}; several previous works have highlighted that interference in accessing shared resources \ab{regulated by bus arbiters and interconnects} is a major source of unpredictability~\cite{RETIS_BANDWIDTH_RESERVATION, RETIS_IS_BUS_ARBITER_FAIR, RETIS_AXI_HYPER_CONNECT, AXIICRT_ARM}.}

In this paper, we present {\axirealm}, an \glsu{axi4}-based, interconnect extension that improves real-time and predictability behavior of \glspl{mcs} by monitoring and controlling \lb{both} the \emph{ingress} and \emph{egress} data streams.
\fr{The architecture is split into two \lbb{sub-systems}: \emph{{\irealm}}, tasked to guard and regulate the \emph{ingress} stream (requests and data issued by managers) with a \emph{budget and time-slicing} approach; \emph{{\erealm}}, tasked to supervise the \emph{egress} stream (data and responses issued by subordinates) with bandwidth and latency statistics, and eventually protect the system from malfunctioning subordinate devices.}

This paper extends our previous work~\cite{benz2023axirealm}, enhancing several aspects:
we provide a more detailed description of the internal \gls{hw} components,
\rev{we include a sequence diagram presenting an example write transaction passing through the various sub-units,}
combine the ingress~\cite{benz2023axirealm} with the egress units~\cite{liang-towards} to \revrep{an}{a} unified {\axirealm} system \rev{capable of shaping the ingress traffic and protecting the interconnect and the system from malfunctioning subordinate devices},
we extend the system-level evaluation by integrating {\axirealm} in an open-source \gls{mcs} characterizing functional, energy, and power performance,\revdel{ and} we extend the \glsu{ip}-level evaluation\revrep{.}{, and we provide a detailed \gls{sota} comparison for the unified system.}
This work provides the following contributions:

\begin{itemize}

    \item \textbf{\axirealm:} We present a scheme to enforce predictable behavior \fr{compatible with any} \gls{axi4}-based interconnect \lbc{which relies on} \emph{observing} and \emph{controlling} both its \emph{ingress} and \emph{egress} data streams \fr{using ad-hoc \gls{hw} methodologies}.
    The resulting architecture \fr{demands minimal \lbc{additional hardware} resources and no \lbc{internal} modifications to the} \lb{baseline} interconnect, enabling portability across diverse \lbc{\gls{soc} targets}.

    \item \textbf{\Gls{hw}-driven traffic controllability:} {\axirealm}'s {\irealm} unit implements a configurable number of subordinate \emph{regions} per manager.
    Each \emph{region} is runtime-programmable with address range, transfer fragmentation size, transfer budget, and reservation period to control the bus traffic through a \emph{time slicing} approach.

    \item \textbf{\Gls{hw}-driven traffic observability:} {\axirealm}'s {\irealm} and {\erealm} units include modules that observe and track per-manager access and interference statistics, such as transaction latency, bandwidth, and interference with each other manager.
    With bandwidth-based observability, {\axirealm} can perform per-manager bandwidth throttling, modulating back-pressure.

    \item \textbf{\Gls{hw}-driven safety measures for malfunctioning subordinates:} We include a \gls{hw} mechanism~\cite{liang-towards} to isolate and reset malfunctioning subordinates individually, taking advantage of {\axirealm}'s {\erealm} latency tracking capabilities to identify response timeouts, mismatching transactions, and invalid handshakes.

    \item  \textbf{IP-level \lbc{characterization}:} We extensively characterize {\axirealm} in a \SI{12}{\nano\metre} technology, presenting an area model as well as timing and latency information.
    
    \item \textbf{In-system \lbb{implementation} assessment:} We evaluate {\axirealm} in an open-source heterogeneous \gls{mcs} research platform~\footnote{\url{github.com/pulp-platform/carfield}}.
    We \lb{demonstrate} the versatility of the proposed approach under interference scenarios between critical and non-critical managers of the system, 
    achieving at least \SI{68}{\percent} of the single-initiator case (over \SI{95}{\percent} when distributing the budget in favor of the \lb{critical manager}).
    \lb{Further, the proposed transfer fragmentation reduces the access latency of the critical manager by \SI{255}{cycles}} \lbb{from 266 to \SI{11}{cycles}.}
    {\axirealm} incurs an area overhead of less than \SI{2}{\percent} in the presented \gls{mcs}.
    
\end{itemize}

The synthesizable and silicon-proven register transfer level description of {\axirealm} and its integration into the presented real-time system are available open-source under a libre Apache-based license~\footnote{\url{github.com/pulp-platform/axi_rt}}.

\lb{This paper is organized as follows.
\Cref{sec:bkgrnd} provides background information.
In \Cref{sec:arch}, we present the architecture of the {\axirealm} system.
\Cref{sec:ip-results} and \Cref{sec:case} present the {IP}-level evaluation of our extension and {\axirealm}'s experimental results integrated into a \gls{mcs}. 
In \cref{sec:relwrk}, we discuss how {\axirealm} compares to \gls{sota}.
Finally, \cref{sec:conclusion} concludes the paper by summarizing its key contributions and achievements.
}

\section{Background}
\label{sec:bkgrnd}

\subsection{The AMBA AXI4 On-Chip Interconnect}

\Gls{axi4} is an industry-standard protocol for high-bandwidth, non-coherent, on-chip communication.
It defines five separate channels for read and write requests (\emph{AR}, \emph{AW}, \emph{W}) and responses (\emph{R}, \emph{B}).
An \gls{axi4} \emph{beat} is the communication in one cycle on an \gls{axi4} channel~\cite{arm_amba_2023}.
An \gls{axi4} \emph{transaction} is the number of beats a manager requires to communicate to a subordinate.
The manager initiates a transaction by emitting an \emph{address and control beat} containing the meta information (address, attributes, and length in beats, ...) over either the \emph{AR} or the \emph{AW} channel.
The \emph{burst} attribute defines the increment mechanism of the write or read addresses during a transaction.

Each \gls{axi4} transaction carries a \emph{\gls{tid}}.
All beats in a transaction must have the same \gls{tid}.
\lbb{The subordinate completes a transaction by sending a response over the \emph{B} channel in the write case or by returning the last read response over the \emph{R} channel} read case.
The protocol also supports \fr{multiple} \emph{outstanding} transactions, i.e., \fr{initiated by the same manager and simultaneously in progress with the same \gls{tid}.} 

Based on \gls{tid}, the protocol dictates three \emph{ordering rules} for write and read transactions.
We recall them in the following:
\circnum{1} for different \glspl{tid}, write data on the \emph{W} channel must follow the same order as the address and control beats on the \emph{AW} channel, as \emph{W} beats do not have a \gls{tid} field;
\circnum{2} transactions with different \glspl{tid} can be completed in any order; on the \emph{R} channel, the read data can be interleaved;
\circnum{3} a manager can have multiple outstanding transactions with the same \gls{tid}, but they must be performed and completed in the order they were requested, for both writes and reads.

This work uses an open-source and silicon-proven implementation of \gls{axi4} network elements~\footnote{\url{github.com/pulp-platform/axi}}.
We refer to a \emph{crossbar} as the main point-to-point network junction between managers and subordinates in the systems (\Cref{sec:arch:ico} and \Cref{sec:case}).

\subsection{MCS Terminology: Essential Insights}

\emph{Criticality} designates \emph{"the level of rigor required to develop safety-critical functions so that the risk of failure can be brought to an acceptable level"}~\cite{MCS_REVIEW_1}.
An \gls{mcs} involves applications with \lbb{different} criticality requirements deployed on the same platform. 
Safety functions in \glspl{mcs} are treated as belonging to the highest safety integrity level unless \emph{independence} between them is guaranteed, i.e., applications achieve freedom from interference with each other.
This implies demonstrating that (i) independence is achieved in both spatial and temporal domains, \todo{and} (ii) violation of independence can be controlled (see \cite{iec61508-3}, Sect. 7.4.2.9).

A way to achieve independence is through \emph{isolation}, or partitioning, of \gls{hw} resources and \gls{sw} components.
Isolation allows the segregation of faults, improves predictability by providing bounds on resource access times~\cite{Cilku2013TowardsTA}, and reduces the \gls{sw} \gls{vv} effort~\cite{MCCU2019}.
\emph{Physical} isolation relies on federated \gls{hw} for each \gls{sw} component.
Hence, resources at all levels are physically decoupled.
\emph{Virtual} isolation establishes partitioned \gls{hw} provisions that allow multiple \gls{sw} components to run on the same \gls{hw} platform, \lbb{namely,} an \gls{mcs}~\cite{MCS_REVIEW_2}.

Within \lbb{\emph{virtual isolation}}, we distinguish between \emph{spatial} and \emph{temporal isolation}.
Spatial isolation means that an application shall not change data used by another application; it can be achieved with virtualization techniques through a \gls{mmu}~\cite{AUTOMOTIVE_PREDICTABLE}.
However, on an integrated \gls{hw} platform, virtual partitions still share resources \lbc{such as caches, interconnects, and memory endpoints,} making their temporal behavior inter-dependent~\cite{Cilku2013TowardsTA, MCS_REVIEW_2}.
Temporal isolation ensures that one application will not cause malfunction of another application by blocking a shared resource over time or consuming another resource execution time.
This can be achieved with adequate scheduling methods in \gls{sw}, or \gls{hw}/\gls{sw} time slicing and fencing~\cite{MCS_REVIEW_1}.

Current industrial and academic activities around \glspl{mcs} do not share a holistic view of fully exploiting \gls{hw}'s potential to isolate executing application layers.
However, some initiatives are being developed and studied~\cite{arm2023mpam}.
A promising direction is that \gls{hw} can cooperate with \gls{sw} by enabling fine-grained \emph{observability} and \emph{controllability} of individual application behavior~\cite{AUTOMOTIVE_PREDICTABLE}.
Proceeding from this premise and terminology background, this work aims to improve the observability and controllability of shared interconnect buses by leveraging time slicing, a temporal isolation technique.

\section{Architecture}
\label{sec:arch}

\begin{figure}[t]
    \centering
    \includegraphics[width=0.95\columnwidth]{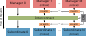}
    \caption{%
        Overview of a generic system extended with {\axirealm}.
        The {\irealm} units monitor and control data from the managers and {\erealm} units guard the subordinate devices.
    }
    \label{fig:realm-system}
\end{figure}

\begin{figure*}[t]
    \centering
    \includegraphics[width=0.9\linewidth]{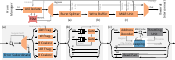}
    \caption{%
        Internals of the {\irealm} unit: (a) \emph{granular burst splitter}, (b) \emph{write buffer}, and (c) \emph{management and regulation unit}.
    }
    \label{fig:irealm-unit}
\end{figure*}

\begin{figure}[ht!]
    \begin{subcaptionblock}{\columnwidth}%
        \centering%
        \includegraphics[width=0.90\columnwidth]{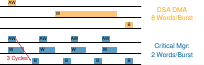}%
        \caption{Isolated waveforms after interconnect.}%
        \vspace{3mm}
        \label{fig:seqiso}%
        \vfill%
        \centering%
        \includegraphics[width=0.90\columnwidth]{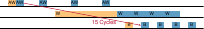}%
        \caption{Unregulated waveform after the interconnect.}%
        \vspace{3mm}
        \label{fig:sequnreg}%
        \vfill%
        \centering%
        \includegraphics[width=0.90\columnwidth]{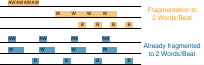}%
        \caption{Regulated waves inside of {\irealm} after the \emph{burst splitter}.}%
        \vspace{3mm}
        \label{fig:seqsplitter}%
        \vfill%
        \centering%
        \includegraphics[width=0.90\columnwidth]{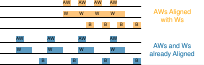}%
        \caption{Regulated waves inside of {\irealm} after the \emph{write buffer}.}%
        \vspace{3mm}
        \label{fig:seqbuffer}%
        \vfill%
        \centering%
        \includegraphics[width=0.90\columnwidth]{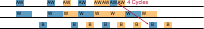}%
        \caption{Regulated waveform after the interconnect.}%
        \vspace{1mm}
        \label{fig:seqrealm}%
        \vfill%
        \centering%
        \includegraphics[width=0.90\columnwidth]{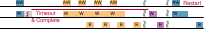}%
        \caption{\rev{Incomplete transaction gets completed by {\erealm} and is restarted.}}%
        \vspace{1mm}
        \label{fig:seqerror}%
    \end{subcaptionblock}
    \caption{Write transaction passing through our {\irealm} unit.}
    \label{fig:seq}
\end{figure}

An overview of a system extended with {\axirealm} is provided in~\cref{fig:realm-system}.
At the \emph{ingress} of the interconnect, \emph{\irealm} units~\cite{benz2023axirealm} monitor and \lbb{shape} traffic injected by managers, enforcing fairness and reducing congestion within the network as well as at the target devices.
The {\irealm} unit tracks the bandwidth and budget on the granularity of a \emph{region}. 
Each region can encompass a subordinate space, combine multiple subordinates, or only cover a fraction thereof.
The number of supported regions can be set \dr{through a SystemVerilog parameter} at design time, and the address space covered by each region through \gls{sw} at runtime.
This is explicitly designed to be independent of the addressing of the interconnect.

At the \emph{egress}, \emph{{\erealm}} units~\cite{liang-towards} guard the subordinate devices. They protect the interconnect and prevent deadline misses of real-time tasks in the case of protocol failures and subordinate regions/devices that extensively delay their responses. %
Our {\erealm} unit provides two messaging options to inform the core of the unresponsive subordinates: interrupts and \gls{axi4} protocol responses.
Moreover, it integrates a framework to isolate, reset, and reinitialize malfunctioning devices within a single cycle~\cite{9567755}.

\subsection{The {\irealm} Unit and Architecture}
\label{sec:arch:irealm}

The {\irealm} unit comprises three main submodules, shown in~\Cref{fig:irealm-unit}: the \emph{burst splitter} (a), the \emph{write buffer} (b), and \gls{mtunit} (c).
At the {\irealm} unit's input, an \gls{axi4} isolation block isolates the manager during dynamic reconfiguration of the unit, once the manager's assigned budget expires, or when commanded through \gls{sw}.

\Cref{fig:seq} shows the function of the {\irealm} unit at two exemplary write transactions from a time-critical manager and a \gls{dma} engine (\Cref{fig:seqiso}).
In the \emph{unregulated} case, \Cref{fig:sequnreg}, the critical transaction experiences a completion latency of up to fifteen cycles.
With our {\irealm} unit activated, the \gls{dma} beats are fragmented after the burst splitter, \Cref{fig:seqsplitter}, and bandwidth reservation is mitigated, \Cref{fig:seqbuffer}, \lbb{by stalling \emph{AWs} until their corresponding \emph{W} beats arrive}.
In this \emph{regulated} case, \Cref{fig:seqrealm}, the transaction latency of the critical manager is at most four cycles.
\rev{\Cref{fig:seqerror} shows the case of a critical subordinate device timing out and how {\erealm} completes and restarts this pending transaction.}

This section explains the architecture and functionality of the {\irealm} unit, detailing how it addresses unfairness from unregulated burst-based communication in \gls{rr} arbitrated interconnect systems. 
It describes how the unit ensures execution predictability using time slicing through static or dynamic budget and period assignments to the managers.

\subsubsection{Granular Burst Splitter}
\label{sec:arch:split}

On-chip interconnects can employ burst-based transactions to increase the efficiency of non-coherent interconnect architectures.
Such transactions increase bus utilization and decrease the addressing overhead.
In heterogeneous \glspl{soc}, transactions of different granularities, e.g., short, cache-line-sized transactions issued by a core and a long burst requested by a \gls{dsa}, are common transaction patterns~\cite{RETIS_AXI_HYPER_CONNECT}.
Classic and fair \gls{rr} arbitration on individual transactions affects bandwidth distribution fairness by increasing the completion latency of short, fine-granular transactions in the presence of long bursts.

As shown in \Cref{fig:irealm-unit}b, the \emph{burst splitter} accepts incoming burst transactions and splits them to a runtime-configurable granularity, from one to 256 beats, according to the \gls{axi4} specification~\cite{arm_amba_2023}.
Any transaction not supported by the burst-splitter is rejected and handled by an \emph{error subordinate}.
For instance, atomic bursts and \emph{non-modifiable} transactions of length sixteen or smaller cannot be fragmented~\cite{arm_amba_2023}.
We store a burst transaction's meta information (address, transaction size, \gls{axi4} flags), emit the corresponding fragmented transactions, and update the address information.
Write responses of the fragmented bursts are coalesced transparently.
Read responses are passed through, except for the \emph{R.last} signal, which is gated according to the length of the original transaction. %
A large granularity requires the write buffer module following the burst-splitter to be large enough to hold a single fragmented write burst.
If a manager only emits single-word transactions, the granular burst splitter can be disabled from the {\irealm} unit to reduce the area footprint. %

\subsubsection{Write Buffer}
\label{sec:arch:buf}

The meta information of a transaction is inherently tied to the data being written.
\Gls{axi4} physically decouples meta information from the payload to increase bus efficiency~\cite{arm_amba_2023}.
However, the write data beats and meta information are not fully decoupled as the write channel does not have a \gls{tid} field.
Most interconnect architectures reserve the bandwidth for an entire write transaction on the \emph{W} channel once the corresponding \emph{AW} is received~\cite{arm_amba_2023}.
\fr{Additionally, according to the \gls{axi4} standard, the \emph{W} channel remains indefinitely allocated to the request's issuer once the request has been propagated through the interconnect.}
\fr{The standard does not specify a maximum delay between the propagation of the request and the provisioning of the corresponding data.}
A manager device can reserve a large transaction by holding the \emph{W} channel, potentially stalling the interconnect by delaying data injection.
In practice, this mechanism is observed with slow manager devices or \gls{dma} units copying data from high-latency or bandwidth-limited endpoints, which cause interference in the downstream memory system, as discussed in~\cite{RETIS_CUT_FORWARD}.
We prevent this behavior by storing the fragmented write burst in a \emph{write buffer}, \Cref{fig:irealm-unit}b.
The buffer forwards the \emph{AW} request and the \emph{W} burst only if the write data is fully contained within the buffer. 
The transaction buffer is configured to hold two \emph{AWs} and one fragmented write burst.

\subsubsection{Monitoring and Regulation Unit}
\label{sec:arch:mr}

In contrast to safety- and time-critical tasks executed on general-purpose processors, \glspl{dsa} often work independently~\cite{RETIS_BANDWIDTH_RESERVATION} and employ double buffering using their large internal memories; this results in memory-intensive phases followed by compute-intensive phases.
The asynchronous nature of \glspl{dsa} accessing the system's memory coupled with coarse-grained synchronization results in unpredictable memory access patterns, increasing the timing uncertainty of critical tasks.

The \gls{mtunit}, presented in \Cref{fig:irealm-unit}c, uses a hardware-implemented period-based bandwidth limiting mechanism to prevent managers from injecting more bandwidth for each subordinate region into the network than allowed.
The \gls{mtunit} is symmetrically designed with identical read and write components. 
Transactions first pass through the address decoder, which maps them to their respective subordinate region. 
A \emph{bus probe} measures the transaction bandwidth and latency, providing this data to the \emph{bookkeeping} unit, which is responsible for budget checks and monitoring bandwidth and latency.
Each {\irealm} unit can track the data sent to each region.

A different time period and budget can be specified for each {\irealm} unit and each subordinate region.
Once activated, this specified budget amount is available and is reduced by every beat passing the unit.
Once one subordinate region's budget exceeds the allocated amount, the number of outstanding transactions is reduced.
The corresponding manager is completely isolated once one budget is depleted using the {\irealm} unit's isolation cell, see \Cref{fig:irealm-unit}.
The budget is automatically renewed once the time period expires.

If the total budget assigned to each manager's {\irealm} unit is less or equal to what the interconnect and the subordinate devices can handle within a period, the {\axirealm} system ensures each manager can use its assigned budget.
The budget distribution should be set according to the real-time task running on the general-purpose cores~\cite{RETIS_BANDWIDTH_RESERVATION}.
The time period, budget checking, and budget renewal are tracked and handled entirely in hardware, allowing the system \lbc{to react} with clock-cycle accuracy.
This allows the {\irealm} unit to set very short periods, ensuring agile regulation and fair bandwidth sharing in the presence of a \gls{dsa} manager.

Furthermore, we extend this monitor to track the average transaction latency issued through the {\irealm} unit.
After a simple profiling run measuring the latency of each manager to each subordinate region in isolation during \gls{vv}, average completion latency \lbc{can be used to reveal} inter-manager interference within the network and its subordinate regions or devices.
\lbc{Thus,} online performance data can help fine-tune the budget and period settings for each manager and assess how well {\irealm} ensures time-critical tasks meet their deadlines.

\begin{figure}[t]
    \centering
    \includegraphics[width=0.87\columnwidth]{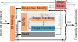}
    \caption{%
        The internal architecture of the {\erealm} unit. The read and write parts are constructed equally.
    }
    \label{fig:erealm-unit}
\end{figure}

\subsection{Interconnect layer}
\label{sec:arch:ico}

We design {\axirealm} to be independent of the system's memory architecture, except for fundamental properties.
Our approach expects the interconnect to ensure a progress guarantee and route transactions using \gls{rr} arbitration, \fr{which is the most common policy in commercial interconnects~\cite{RETIS_AXI_HYPER_CONNECT, AXIICRT_ARM, BSC_ARM_QOS400}}.
{\axirealm} is primarily intended for mixed-critical systems where \glspl{dsa} require a high-performance interconnect to satisfy their data demand, and critical actors rely on real-time guarantees.
We specifically choose an \gls{rr}-arbitration mechanism over more classical real-time distribution patterns, like \gls{tdma}\rev{~\cite{poletti2003performanceanal}}, to maintain the high-performance memory access required by the \rev{\glspl{dsa} in today's emerging heterogeneous} \revrep{system's DSA}{systems}.
We aim to enhance the determinism and fairness of a classic \gls{rr}-arbitrated interconnect by minimally intruding on its design, using lightweight helper modules at its boundaries.

\rev{Thanks to the independence from instance-specific assumptions on the interconnect architecture and to the compliance with the \gls{axi4} specification of our {\axirealm} architecture, verification and maintenance are facilitated by allowing the use of unit-level standalone verification infrastructure.}

As mentioned in~\cref{sec:bkgrnd}, the in-system case study and \gls{ip}-level evaluation presented in this work use a point-to-point-based interconnect constructed from \gls{axi4} crossbars.
{\axirealm} can handle hierarchical point-to-point interconnects thanks to the concept of subordinate regions.
\rev{While outside of the scope of this work, we also ensured compatibility of {\axirealm} with \gls{axi4}-based \gls{noc} architectures, such as presented in~\cite{fischer2024floonoc645gbpslink015}}.

\subsection{The {\erealm} Unit and Architecture}
\label{sec:arch:erealm}

The {\erealm} unit uses a \gls{hw}-based approach to monitor the latency of transactions sent to a subordinate device or region, responding promptly to unexpected misbehavior or malfunctions that disrupt the predictability of real-time tasks. 
This unit is essential because the assumption of perfect behavior by subordinate devices, often made in \gls{sota}~\cite{RETIS_IS_BUS_ARBITER_FAIR, RETIS_AXI_HYPER_CONNECT}, does not hold in real-world scenarios.
Additionally, the {\erealm} proactively ensures protocol compliance for all transactions without impacting system throughput or latency.

When responses from a subordinate device exceed user-programmable timeouts or when a protocol violation is detected, {\erealm} completes the outstanding transactions and communicates the cause of the issue with the core either using interrupts or the \gls{axi4} response channel.
Error information, including \gls{tid}, address, and the \emph{specific transaction stage} in which the error occurred, are logged into registers.
The unit can reset the connected subordinates through an agile reset controller~\cite{9567755} either within one clock cycle upon fault detection, or when commanded by the core as part of the fault handling. 
Overall, {\erealm} guards subordinate devices, guaranteeing responses within user-defined time frames, preventing the interconnect from locking up.

The architecture of {\erealm} is shown in \Cref{fig:erealm-unit}.
An ID remapper at the unit's input compacts the typically sparsely used \gls{tid} space, requiring fewer \gls{tid} bits to track all transactions.
The data path is then split into a similarly constructed \emph{Write} and a \emph{Read module}, presented in more detail below.

\subsubsection{Dynamic Outstanding Transaction Queue (DOTQ)}

{\axirealm} supports multiple outstanding,  multi-id transactions commonly occurring when accessing high-performance subordinate devices; requiring dynamic tracking of multiple data streams, each with several outstanding transactions.

The {\erealm} unit manages this through a dynamic queue consisting of three linked tables: an \emph{ID Head-Tail (HT)} table, a \emph{Transaction Linked Data (LD)} table, and a \emph{Write (W)} or \emph{Read (R)} table present in the write and read path, respectively.
The HT table keeps track of active \glspl{tid} and enforces ordering for transactions with the same \gls{tid} and supports efficient \gls{tid} lookups without scanning through all transactions in the LD table.
The LD table stores metadata such as \gls{tid}, address, state, latency, and timeout, allowing detailed tracking of each outstanding transaction.
Finally, the W/R table ensures that write data maintains the correct sequence with address beats, aligning data properly even when \glspl{tid} are not explicitly available on the write data channel.
The tracking capacity is defined by two design parameters: the maximum number of unique \glspl{tid} and the maximum number of transactions each unique \gls{tid} can support simultaneously.

\subsubsection{Stage-Level Tracking for each Transaction}
\label{sec:arch:erealm:stage}

\begin{figure}[t]
    \centering
    \includegraphics[width=0.95\columnwidth]{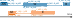}
    \caption{%
        Tracked stages in the {\erealm} unit.
    }
    \label{fig:erealm-stage}
\end{figure}

\Gls{axi4} transactions occur in multiple \emph{stages}: address and meta information is sent first, followed by the data beats, and finally, the response stage.
As shown in \Cref{fig:erealm-stage}, the counter-based tracking logic monitors \emph{six} and \emph{four} stages for write and read transactions, respectively. 
The first stage is the initial handshake transferring address and meta information \emph{(ax\_valid to ax\_ready)} to confirm the transaction acceptance.
For the write case, this is followed by the transition from address acceptance to data availability \emph{(aw\_ready to w\_valid)} to ensure that data beats readiness promptly follows.
Monitoring continues with the acceptance of the first data beat \emph{(w\_valid to w\_ready)} and in a subsequent stage, the data beat from the first to the last beat \emph{(w\_valid to w\_last)} to guarantee continuous and correct data flow.
After the last write data beat, the monitoring tracks from \emph{(w\_last to b\_valid)}, which is to confirm that the subordinate device sends the write response in time.
Finally, the transition from write response valid to response ready \emph{(b\_valid to b\_ready)} checks that the acknowledgment is properly issued by the manager device, which marks the end of the transaction. 
For read operations, the transition from address acceptance to data availability \emph{(rw\_ready to r\_valid)} is monitored, along with the data beats \emph{(r\_valid to r\_last)} and the timely delivery of the read response.

\subsubsection{\revrep{Dynamic Time}{Stage} Budget Allocation}

In \gls{axi4} systems, transactions of the same \gls{tid} are processed sequentially.
The \revrep{time}{stage} budget for each transaction is dynamically \revrep{scaled}{calculated} \revrep{based on its}{by multiplying the per-word stage budget, obtained by profiling the system during \gls{vv}, with the} burst length\revrep{ and the cumulative burst lengths of prior transactions recorded in the transaction LD table.}{, granting longer bursts more time to complete.}

For write transactions, this impacts the time budget from the \emph{aw\_ready to w\_first} stage, while the time from \emph{w\_first to w\_last} is counted only when the transaction begins servicing, thus excluding prior transaction latency.
A similar dynamic \rev{stage} budgeting method is applied for read transactions.
This ensures that transactions, particularly those with large data beats, have sufficient time to complete.

\subsection{\lbc{Configuration} Interface}
\label{sec:arch:prog}

In its basic configuration, both {\irealm} and {\erealm} units are configured through a shared set of memory-mapped registers, as shown in \Cref{fig:realm-system}.
The shared configuration register file can be physically decoupled to increase the scalability of the {\axirealm} architecture in larger designs.
Configurations with dedicated configuration register files for each {\irealm} and {\erealm} unit are supported.

The register values are reset to a default configuration on startup, the {\erealm} units are deactivated, and the {\irealm} are bypassed.
In this reset state, the {\axirealm} system is inert, interconnect accesses are unregulated, and no additional latency is introduced.
One privileged manager, e.g., the booting core from a secure domain, programs the {\axirealm} system.
The system's memory protection or {\axirealm}'s \emph{bus guard} restricts configuration space access to privileged managers.
{\axirealm} can be dynamically reconfigured during the system's runtime.

\begin{figure}[t]
    \centering
    \includegraphics[width=0.82\columnwidth]{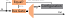}
    \caption{%
        The architecture of the \emph{bus guard}.
    }
    \label{fig:bus-guard}
\end{figure}

\subsection{Bus Guard}
\label{sec:arch:bus-guard}

{\axirealm} 's configuration space must be protected against malicious or erroneous accesses.
Most systems use physical memory protection units or virtual memory space, e.g., through \glspl{mmu}, to isolate critical configuration spaces.

\lbc{Even in} systems with no such protection device, our minimal \emph{bus guard} unit, presented in \Cref{fig:bus-guard}, restricts unwanted access to the configuration interface. %
After a system reset, a trusted manager must claim ownership of the configuration space by writing to a \emph{guard register} within the \emph{bus guard}.
In the unclaimed state, every access to the configuration space except for the \emph{guard register} returns an error. %
Once a manager has claimed the address space, it can perform a \emph{handover} operation to transfer the exclusive read/write ownership to any other manager in the system.
The \emph{bus guard} differentiates between managers using their unique \gls{tid}. %

\section{IP-level \lbc{Characterization}}
\label{sec:ip-results}

\newcommand{\rot}[1]{\rotatebox[origin=c]{90}{#1}}
\newcommand{\tilt}[1]{\hspace{-1cm}\rotatebox[origin=c]{32}{#1}}
\newcommand{\noc}{\textcolor{ieee-dark-grey-40}{0}}
\newcommand{\dk}{\textcolor{ieee-bright-red-100}{?}}
\newcommand{\na}{\textcolor{ieee-dark-grey-40}{N.A.}}
\newcommand{\nad}{\textcolor{ieee-dark-grey-40}{-}}
\newcommand{\dl}[2]{\makecell[cc]{#1 \\ #2}}
\newcommand{\dll}[2]{\makecell[cl]{#1 \\ #2}}
\newcommand{\dlb}[2]{\makecell[cc]{\textbf{#1} \\ \textbf{#2}}}
\newcommand{\tl}[3]{\makecell[cc]{#1 \\ #2 \\ #3}}
\newcommand{\tll}[3]{\makecell[cl]{#1 \\ #2 \\ #3}}
\newcommand{\tlb}[3]{\makecell[cc]{\textbf{#1} \\ \textbf{#2} \\ \textbf{#3}}}

\begin{table}[t]
    \centering
        \centering
        \caption{%
            Area contribution \emph{weights} of {\axirealm}'s building blocks as a function of their parameters. %
            All numbers are in \si{\GE}, at \SI{1}{\giga\hertz} using typical conditions.%
        }%
        \label{tab:ooc-area}
        \centering
        \resizebox{1\columnwidth}{!}{%
            \begin{threeparttable}
                \renewcommand{\arraystretch}{1.1}
                \renewcommand{\tabcolsep}{2pt}
                \begin{tabular}{@{}lllccccccccc@{}}
                    \toprule
                    \multirow{7}{*}{\rot{Config. Registers}} &
                    \multirow{3}{*}{PUR} &
                    (i) Status &
                    \noc &
                    \noc &
                    \noc &
                    \noc &
                    \noc &
                    \noc &
                    \noc &
                    \noc &
                    24.6 \\
                    &
                    &
                    (i) Budget/Period &
                    \noc &
                    \noc &
                    \noc &
                    \noc &
                    \noc &
                    \noc &
                    \noc &
                    \noc &
                    1320 \\
                    &
                    &
                    (i) Region Bound. &
                    20.6 &
                    \noc &
                    \noc &
                    \noc &
                    \noc &
                    \noc &
                    \noc &
                    \noc &
                    \noc \\
                    \arrayrulecolor{ieee-dark-black-40}\cdashline{2-12}
                    &
                    \multirow{3}{*}{PU} &
                    (i) Config &
                    \noc &
                    \noc &
                    \noc &
                    \noc &
                    \noc &
                    \noc &
                    \noc &
                    \noc &
                    83.5 \\
                    &
                    &
                    (e) Status/Config &
                    \noc &
                    \noc &
                    \noc &
                    \noc &
                    \noc &
                    \noc &
                    \noc &
                    \noc &
                    9.7 \\
                    &
                    &
                    (e) R/W Budget &
                    \noc &
                    \noc &
                    \noc &
                    \noc &
                    \noc &
                    \noc &
                    770 &
                    \noc &
                    \noc \\
                    \arrayrulecolor{ieee-dark-black-40}\cdashline{2-12}
                    &
                    \multirow{1}{*}{PS} &
                    Bus Guard &
                    \noc &
                    \noc &
                    \noc &
                    \noc &
                    \noc &
                    \noc &
                    \noc &
                    \noc &
                    261 \\
                    \arrayrulecolor{ieee-dark-black-100}\midrule
                    &
                    \multirow{2}{*}{PUR} &
                    Tracking Cnts. &
                    \noc &
                    \noc &
                    \noc &
                    \noc &
                    \noc &
                    \noc &
                    \noc &
                    \noc &
                    1930 \\
                    &
                    &
                    Region Decoders &
                    20.8 &
                    \noc &
                    \noc &
                    \noc &
                    \noc &
                    \noc &
                    \noc &
                    \noc &
                    \noc \\
                    \arrayrulecolor{ieee-dark-black-40}\cdashline{2-12}
                    \multirow{6}{*}{\rot{\irealm}} &
                    \multirow{4}{*}{PU} &
                    Isolate/Throttle &
                    3.5 &
                    2.7 &
                    9.0 &
                    \noc &
                    \noc &
                    \noc &
                    \noc &
                    \noc &
                    267 \\
                    &
                    &
                    Burst Splitter &
                    49.3 &
                    1.5 &
                    729 &
                    \noc &
                    \noc &
                    \noc &
                    \noc &
                    \noc &
                    4840 \\
                    &
                    &
                    Meta Buffer &
                    38.1 &
                    \noc &
                    \noc &
                    \noc &
                    \noc &
                    \noc &
                    \noc &
                    \noc &
                    1310 \\
                    &
                    &
                    Write Buffer &
                    \noc &
                    \noc &
                    \noc &
                    \noc &
                    \noc &
                    264 &
                    \noc &
                    \noc &
                    11.4 \\
                    \arrayrulecolor{ieee-dark-black-100}\midrule
                    \multirow{6}{*}{\rot{\erealm}} &
                    \multirow{6}{*}{PU} &
                    ID Remap. &
                    \noc &
                    \noc &
                    \noc &
                    \noc &
                    \noc &
                    \noc &
                    \noc &
                    \noc &
                    \noc \\
                    &
                    &
                    Stage Cnts. &
                    \noc &
                    \noc &
                    \noc &
                    \noc &
                    \noc &
                    \noc &
                    \noc &
                    129 &
                    735 \\
                    &
                    &
                    HT Table &
                    \noc &
                    \noc &
                    201 &
                    \noc &
                    \noc &
                    \noc &
                    \noc &
                    \noc &
                    \noc \\
                    &
                    &
                    LD Table &
                    \noc &
                    \noc &
                    \noc &
                    51 &
                    \noc &
                    \noc &
                    \noc &
                    \noc &
                    \noc \\
                    &
                    &
                    R/W Table/Ctrl. &
                    \noc &
                    \noc &
                    \noc &
                    \noc &
                    \noc &
                    \noc &
                    329 &
                    \noc &
                    356 \\
                    &
                    &
                    Reset Ctrl. &
                    \noc &
                    \noc &
                    \noc &
                    \noc &
                    \noc &
                    \noc &
                    \noc &
                    \noc &
                    1270 \\
                    \arrayrulecolor{ieee-dark-black-100}\midrule
                    &
                    &
                    &
                    \tilt{Addr Width~\tnote{a}~\tnote{b}}             &
                    \tilt{Data Width~\tnote{a}~\tnote{b}}             &
                    \tilt{Num Pending~\tnote{c}}                      &
                    \tilt{Num \glspl{tid}}                            &
                    \tilt{Buffer Depth~\tnote{c}}                     &
                    \tilt{Storage Size~\tnote{a}~\tnote{d}~\tnote{e}} &
                    \tilt{Num Counters~\tnote{i}}                               &
                    \tilt{Counter Storage~\tnote{j}}                            &
                    \tilt{Constant~\tnote{f}}                         \\
                    \arrayrulecolor{ieee-dark-black-100}\bottomrule
                \end{tabular}

                \begin{tablenotes}[para, flushleft]
                    \fontsize{7.7pt}{7.7pt}\selectfont
                    \item[a] In [bit]
                    \item[b] Evaluated \SIrange{32}{64}{\bit}
                    \item[c] Evaluated \SIrange{2}{16}{elements}
                    \item[d] Product of \emph{Buffer Depth} and \emph{Data Width}
                    \item[e] Evaluated \SIrange{256}{8192}{\bit}
                    \item[f] Base area independent of params
                    \item[i] Product of \emph{Num Pending} and\emph{Num \glspl{tid}}
                    \item[j] Product of \emph{Num Counters} and the counter width (\SIrange{10}{32}{\bit})
                \end{tablenotes}
            \end{threeparttable}
        }
\end{table}

This section provides an extensive area, timing, and latency model to enable quick design-space exploration and promote fair comparison with other works.
For gate-level assessment, we use {\gfs} {\gftech} node with a 13-metal stack and 7.5-track standard cell library in the typical corner. %
We synthesize the designs using {\dc} in topological mode to account for place-and-route constraints, congestion, and physical phenomena.
We provide all area results in \emph{gate equivalent (\si{\GE})}, a technology-independent circuit complexity metric\rev{, allowing comparisons among technology nodes}.
A \si{\GE} represents the area of a two-input, minimum-strength {NAND} gate.

\subsection{Area Model}
\label{sec:ooc:area}

Our linear area model, given in \Cref{tab:ooc-area}, allows us to estimate a system's {\axirealm} configuration given the number of {\irealm} and {\erealm} units with their respective parameters.
We synthesize our {\axirealm} system to construct the model, sweeping the parameter space.
The resulting areas are correlated with the input parameters, and a linear model is fitted. 
The model is divided into three categories: \emph{Configuration Registers}, \emph{\irealm}, and \emph{\erealm}.
Each category is grouped into the sub-categories: \emph{\gls{ps}}, \emph{\gls{pu}}, and \emph{\gls{pur}}.
To estimate the area of an {\axirealm} system or its components, the system's desired configuration is determined. 
This includes the number of {\irealm} and {\erealm} units, configuration register files, and regions, as well as the \gls{ip}'s desired SystemVerilog parameters.
The area of the individual sub-units is given by a linear function of the \gls{ip} parameters with the coefficient in \Cref{tab:ooc-area}.
The total {\axirealm} area can be \lbc{obtained by summing over} the individual sub-unit's contributions.

\lbb{%
We use the {\axirealm} \gls{mcs}'s configuration (\Cref{tab:params}) presented in \Cref{sec:case} as an example of how to use our model.
The \emph{bus guard} is the only \gls{ps} item with a constant contribution of \SI{261}{\GE}.
For each \gls{pu} and \gls{pur} elements we evaluate
\setlength{\belowdisplayskip}{6pt} \setlength{\belowdisplayshortskip}{6pt}
\setlength{\abovedisplayskip}{6pt} \setlength{\abovedisplayshortskip}{6pt}
\begin{equation*}
A_{contrib} = \sum_{i}{param_i}*{weight_i} + constant
\end{equation*}
to obtain their respective area contribution.
E.g., for the \emph{write buffer}, we multiply \SI{264}{\GE} with the storage size of 256 and add \SI{11.4}{\GE}.
Our example features three {\irealm} units, bringing the total area contribution of all \emph{write buffers} in the {\axirealm} system to \SI{203}{\kGE}. 
The total modeled area, presented in \Cref{tab:params}, is calculated by summing all \gls{ps}, \gls{pu}, and \gls{pur} contributions together. %
}

\rev{Our {\axirealm} architecture does not have inherent limitations in terms of throughput, the supported amount of \glspl{tid}, and the number of outstanding transfers as long as the units are tuned to the encompassing system and its use cases.}

\begin{table}
    \centering
    \caption{Parametrization, the resulting modeled, and actual area of the {\irealm} and {\erealm} units in {\carfield} (\Cref{sec:case}).}
    \resizebox{\columnwidth}{!}{%
        \begin{threeparttable}
            \renewcommand{\arraystretch}{1.2}
            \renewcommand{\tabcolsep}{3pt}
            \begin{tabular}{@{}llllccccccccc@{}}
                \toprule
                \dll{\textbf{Sub-}}{\textbf{system}} &
                \dll{\textbf{Num.}}{\textbf{Units}} &
                \dll{\textbf{Num.}}{\textbf{Regions}} &
                \multicolumn{8}{l}{\dll{\textbf{SystemVerilog}}{\textbf{Parameters}}} &
                \lbb{\tll{\textbf{Model}}{\textbf{Area}}{\textbf{[\si{\kGE}]}}} &
                \tll{\textbf{Design}}{\textbf{Area}}{\textbf{[\si{\kGE}]}} \\
                \midrule
                \emph{\irealm} &
                3 &
                2 &
                \SI{48}{\bit} &
                \SI{64}{\bit} &
                16 &
                \nad &
                4 &
                \SI{256}{\bit} &
                \nad &
                \nad &
                330 &
                328 \\
                \emph{\erealm} &
                1 &
                \nad &
                48~\tnote{a} &
                64~\tnote{a} &
                2 &
                2 &
                \nad &
                \nad &
                20 &
                \SI{200}{\bit} &
                50 &
                45 \\
                \midrule
                &
                &
                &
                \tilt{Addr Width}           &
                \tilt{Data Width}           &
                \tilt{Num Pending}          &
                \tilt{Num \glspl{tid}}      &
                \tilt{Buffer Depth}         &
                \tilt{Storage Size}         &
                \tilt{Num Counters}         &
                \tilt{Counter Storage}      &
                \\
                \bottomrule
            \end{tabular}
        \end{threeparttable}
        \label{tab:params}
    }
\end{table}

\subsection{Timing and Latency}
\label{sec:ooc:timing}

The {\axirealm} architecture and its units are designed to achieve clock speeds exceeding \SI{1.5}{\giga\hertz} (corresponding to \SI{25}{logic} levels) in {\gftech} when combined with optimized \gls{axi4} \glspl{ip} for \glspl{asic}.
The achieved frequency can be further increased at the cost of additional latency by either adding \gls{axi4} cuts around the {\axirealm} units or introducing pipelining into the {\axirealm} units.
The {\irealm} unit adds no additional cycle of latency when bypassed and introduces one cycle through the write buffer (\Cref{sec:arch:buf}) when active. 
The {\erealm} unit adds no additional latency, whether bypassed or active.

\section{Case Study: Open-Source Automotive MCS}
\label{sec:case}

\emph{{\carfield}} establishes a heterogeneous platform for mixed-criticality systems and application \lbc{across} domains like automotive, space, and cyber-physical embedded systems.
At the core of {\carfield}, the \emph{host} domain consists of a Linux-capable dual-core \emph{CVA6} system enhanced with virtualization extensions, namely {\riscv}'s \emph{H-extension} and virtualized fast interrupts.
The platform is complemented by a \emph{safety} and a \emph{security} domain, allowing for reliable operation and secure boot, respectively.
{\carfield} computational capabilities are enhanced through general-purpose \glspl{dsa}.
\lbc{We instantiate two \glspl{dsa}:} one specialized for integer and one for floating-point workloads.
Each accelerator features \lbb{an internal} \gls{dma} engine to copy data between its private \gls{spm} and the \gls{mcs}'s main storage. 
{\carfield} features two memory endpoints, a \SI{512}{\kibi\byte} banked \gls{l2} memory and an off-chip {DRAM} accessed through a \gls{llc}.
Each rank of the platform's \gls{llc} can be configured either as software-managed \gls{spm} or cache for the {DRAM}. 
All five domains are connected through a 64-bit point-to-point \gls{axi4} crossbar.%

\begin{figure}[t]
    \centering
    \includegraphics[width=0.9\columnwidth]{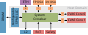}
    \caption{%
        Architectural block diagram of the {\carfield} platform.
    }
    \label{fig:carfield-bd}
\end{figure}

We integrate {\axirealm} into {\carfield} by adding four {\irealm} units at the crossbar's ingress of all time-critical managers the two \glspl{dsa} and the two CVA6 cores.
An {\erealm} unit ensures real-time responses from the platform's \gls{eth} controller, which accesses time-critical sensor data. 
\lbc{Characterization in {\carfield} shows \gls{eth} to be prone to data loss, and it further exhibits strong fluctuations in timing behavior\revrep{.}{, thus being a representative candidate for a subordinate guarded by {\erealm}}.}
\Cref{fig:carfield-bd} shows the enhanced architectural block diagram enhanced with our {\axirealm} units.

\begin{figure}[t]
    \centering
    \includegraphics[width=\columnwidth]{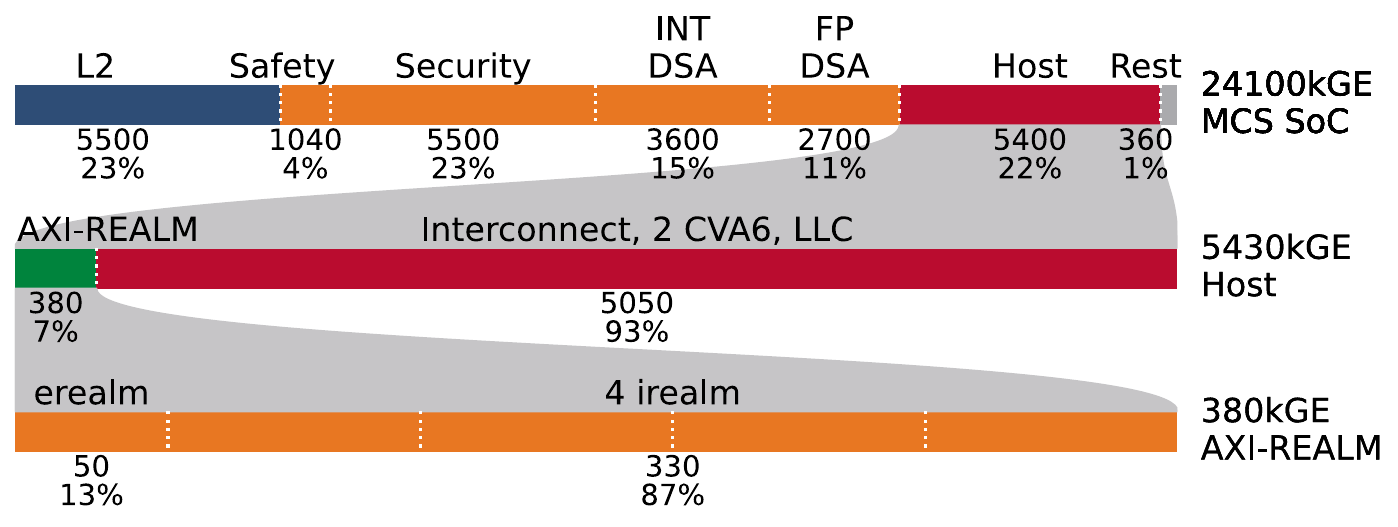}
    \caption{%
        {\carfield}'s hierarchical area including {\axirealm}.
    }
    \label{fig:carfield-area}
\end{figure}

\begin{figure}[t]
    \centering
    \includegraphics[width=\columnwidth]{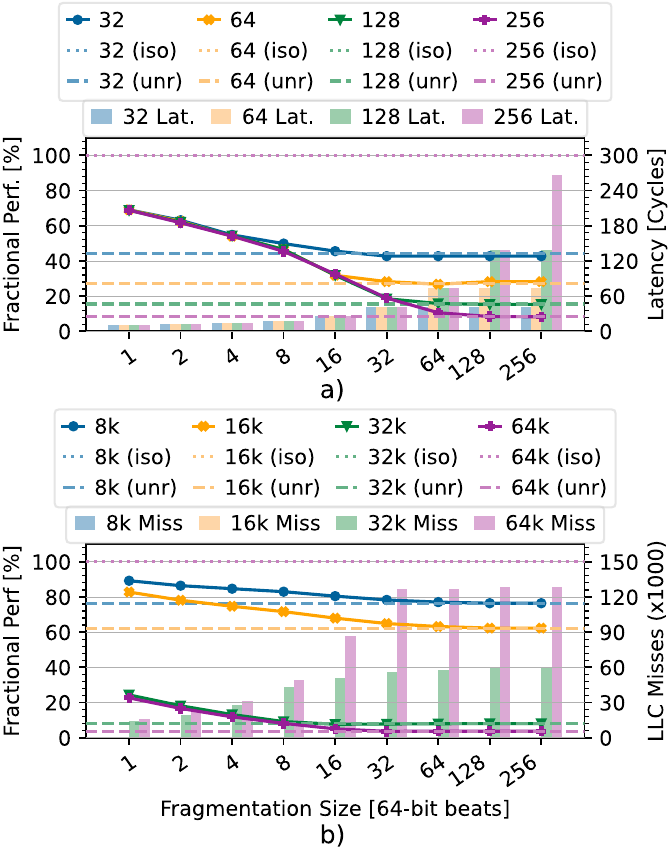}
    \caption{%
        Performance results of CVA6 copying data between (a) \gls{spm} and \gls{l2} with the \gls{dsa} accessing \gls{spm} at various granularities in 64-\si{bit} \si{beats} and (b) {DRAM} and \gls{l2} with the \gls{dsa} accessing {DRAM} at various problem sizes in \si{byte}.
        \dr{\emph{iso} denotes the isolated, \emph{unr} the unregulated performance.} 
    }
    \label{fig:result:memcpy}
\end{figure}

\subsection{Area Impact}
\label{sec:case:area}

We synthesize {\carfield} with the {\axirealm} extensions in {\gfs} \SI{12}{\nano\metre} node using typical timing corners.
\Cref{fig:carfield-area} presents the total \gls{soc} area of \SI{24}{\mega\GE} and the hierarchical area contributions of our units introduced.
The {\irealm} units incur \lbb{a total} overhead of \SI{330}{\kilo\GE}, contributing \SI{1.4}{\percent} to the total area.
The {\erealm} unit uses \SI{50}{\kilo\GE} (\SI{0.21}{\percent}) of the \gls{soc}'s area.
The parameterization of the {\axirealm} units implemented in {\carfield} is given in \Cref{tab:params}, \Cref{sec:ip-results}.

\subsection{Synthetic Performance Analysis of {\irealm}}
\label{sec:case:synth-eval}

We first evaluate the functional performance of the {\irealm} architecture using a memory-bound synthetic benchmark, which emulates the real-time-critical task, to maximize the effects of interconnect and subordinate interference between the processor cores and the platform's \glspl{dsa}.
This synthetic benchmark uses {CVA6} to copy data between different memory locations while a \gls{dma} engine in one of the \glspl{dsa} performs data transfer operations.
The default configuration is to copy \SI{1}{\kibi\byte} of data with the core from {\carfield}'s hybrid \gls{llc}, configured as a \gls{spm}, to \gls{l2} memory while the \gls{dsa} causes interference in the \gls{spm} using long bursts of 256 beats.
Large and equal budget periods as well as a fragmentation size of one are used if nothing else is specified.
Results from application benchmarks are presented in \Cref{sec:case:bmk-eval}.

\begin{figure*}[t]
    \centering
    \includegraphics[width=\linewidth]{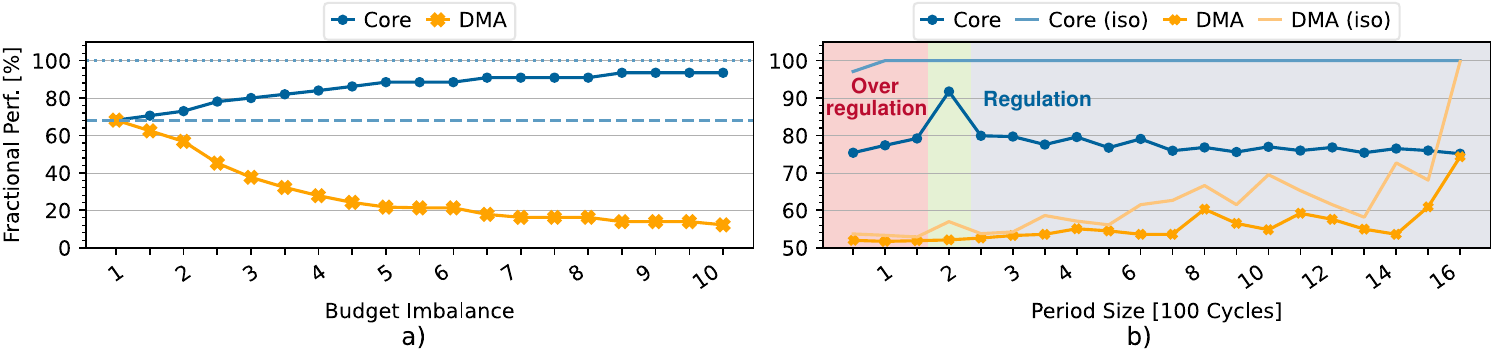}
        \caption{%
           (a) Fractional performance at different budget imbalances favoring the critical manager assuming fragmentation one and (b) fractional performance at different period sizes assuming fragmentation size one and equal budget.
        }
    \label{fig:period-bmk-perf}
\end{figure*}

\subsubsection{Controlling Fairness: Burst Fragmentation}
\label{sec:case:frag}

We configure the {\irealm} units of {CVA6} and one of the \gls{dsa} to fragment transactions at different granularities without any budget limitation.
For this synthetic assessment, the write buffer is disabled, as both managers are under our control, eliminating the possibility of bandwidth stealing.
To simulate different \gls{dsa} workloads, we vary the \gls{dma} transaction length in \Cref{fig:result:memcpy}a from 32 to 256 64-\dr{\si{bit}} words.
Fragmenting all beats to single-world granularity results in the best performance independent of the nature of the \gls{dma} transfer; in this setting, {CVA6} achieves \SI{68}{\percent} of its isolated performance.
The worst slowdown can be observed in the presence of 256-word-long bursts without fragmentation activated, which represents the unregulated case: {CVA6} only achieves \SI{1}{\percent} of the isolated performance.
\lbb{The latency of each core access increases from \SI{11}{cycles} to \SI{266}{cycles}, as core transfers are interleaved by the 256-\si{cycle}-long \gls{dsa} transfers}.

In a second experiment, \lbb{\Cref{fig:result:memcpy}b}, we copy data between the system's external DRAM cached by the \gls{llc} and the \gls{l2} memory.
The \gls{dsa} \gls{dma} is configured to emit 256-word-long bursts at varying data set sizes.
Due to conflict misses between the \gls{dsa} and the core, and capacity misses when the \gls{dsa} transfer size surpasses the \gls{llc} capacity, \lbb{the fraction of isolated performance} drops from \SI{80}{\percent} (\SI{16}{\kilo\byte}) to \SI{23}{\percent} (\SI{64}{\kilo\byte}) at single-word fragmentation.
{\axirealm} can help mitigate interference in accessing a shared cached memory location, allowing {\carfield} to achieve up to \SI{23}{\percent} of the isolated performance \lbb{as opposed to \SI{4}{\percent} without regulation}.
To improve performance further, complementary regulation strategies must be \lbb{put in place for the shared} \gls{llc}. 
For example, cache coloring or partitioning can mitigate conflict misses or \gls{dma} cache bypassing to eliminate capacity misses.

\subsubsection{Period and Budget Considerations}
\label{sec:case:per-bud}

\begin{figure}[t]
    \centering
    \includegraphics[width=\columnwidth]{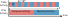}
        \caption{%
            Schedule of the periodic (\emph{p}) real-time-critical transactions running on {CVA6} and the data copy operation (period \emph{P}) by the \gls{dsa} \gls{dma}.
        }
    \label{fig:period-bmk-plot}
\end{figure}

This section assesses {\axirealm}'s period and budget functionality at fixed fragmentation. 
In particular, we demonstrate that tuning each manager's period and budget can prioritize traffic of certain managers over others and even increase fairness~\cite{RETIS_BANDWIDTH_RESERVATION} further compared to solely acting on the fragmentation size.
We observe that the budget imbalance favoring the real-time task restores performance up to \SI{95}{\percent} of the isolated case, at a performance detriment to the \gls{dsa} \gls{dma}, see \Cref{fig:period-bmk-perf}a.
Similarly, the period can be used as a knob for online traffic regulation~\cite{RETIS_BANDWIDTH_RESERVATION}.
{\axirealm} does not limit the managers in how to spend the budget within each period.
Larger periods introduce less regulation overhead but allow \gls{dsa} managers to cause more interference.
\lbb{%
We assume the periodic execution schedule for the critical manager and the \gls{dma} \gls{dsa} given in \Cref{fig:period-bmk-plot}.
The first has a period \emph{p}, \SI{200}{cycles}, and the latter a period \emph{P}, set to \SI{1600}{cycles}.
Both managers utilize the interconnect \SI{50}{\percent} within their period; for the critical manager this corresponds to \SI{800}{\byte} in \SI{200}{cycles}, for the \gls{dma} \SI{6400}{\byte} in \SI{1600}{cycles}.
The corresponding {\irealm} units are configured equally to a fragmentation size of one.
The regulation time period is swept from 50 to \SI{1600}{cycles} with the budget set to half the maximum transfer size possible during the regulation period. 
E.g., the budget for the critical manager and the \gls{dma} are set to \SI{6400}{\byte} each when selecting a period of \SI{1600}{cycles}. 
}

\Cref{fig:period-bmk-perf}b shows the fractional performance of the \gls{dma} and the core given selected period sizes.
The performance of the \gls{dma} decreases when the {\irealm} period falls below the \gls{dma}'s period of \emph{P}, regardless of whether the critical manager is active.
This happens due to \emph{overregulation} as every \gls{dma} transaction no longer fits a period, interrupting it at least once. %

The critical manager nearly matches isolated memory performance --- over \SI{93}{\percent} of the isolated case --- when the {\irealm} period aligns with the core's task period of \emph{p}.
Below this critical period, the core's transfer is again overregulated.

\subsubsection{Power and Energy Efficiency Analysis}
\label{sec:case:power}

\begin{figure}[t]
    \centering
    \includegraphics[width=\columnwidth]{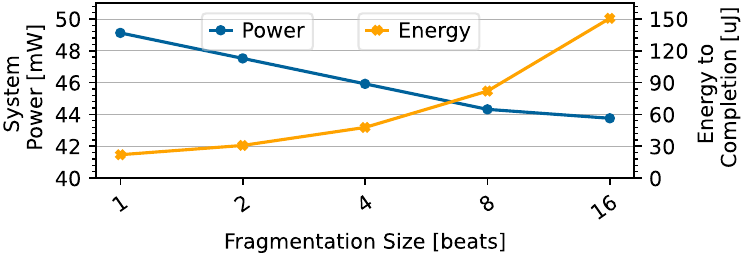}
    \caption{%
        Power and energy of CVA6 copying data between \gls{spm} and \gls{l2} with the \gls{dsa} accessing \gls{spm} given different fragmentation sizes.%
    }
    \label{fig:carfield-power}
\end{figure}

\Cref{sec:case:frag} establishes full fragmentation as the configuration achieving the best performance in the presence of \gls{dsa} interference.
However, fragmenting transfers to word-level accesses increases the switching activity (e.g., the address changes on every access) and, thus, the power consumed by the interconnect and the subordinate devices.
We evaluate the energy and power consumption of our synthetic benchmark running on {\carfield} using timing-annotated switching activity on a post-layout netlist using {\pt}.
We vary the fragmentation size of the {\irealm} units from one to sixteen beats.
The peak power consumed by the host domain, which includes the host, the \gls{mcs}'s main interconnect, and the \gls{spm} memory, is linearly increasing with decreasing fragmentation size.
When evaluating {CVA6}'s energy spent to copy \SI{1}{\kibi\byte} of data, the energy required is minimal at a fragmentation size of one.
Even though the activity of fragmenting transactions is increasing the power consumption of the benchmark, the reduction in execution time outweighs the increase in activity, \rev{as seen in \Cref{fig:carfield-power}}.

\subsection{Synthetic Performance Analysis of {\erealm}}
\label{sec:case:synth-eval-erealm}

\begin{figure}[t]
    \centering
    \includegraphics[width=\columnwidth]{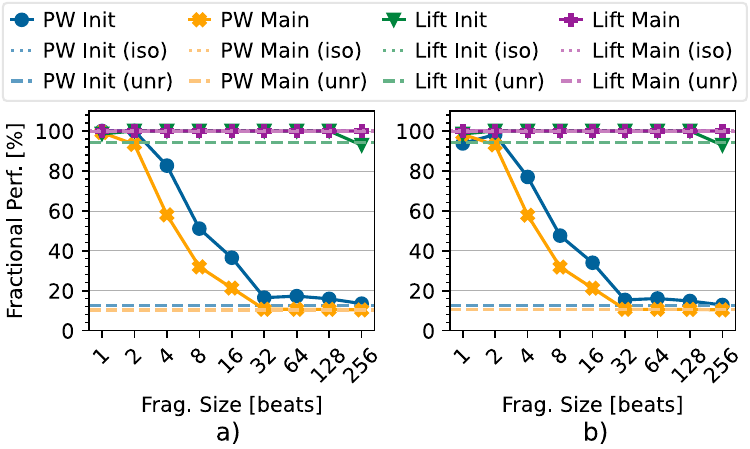}
    \caption{%
        TACLeBench~\cite{TACLeBench} performance of {CVA6}
        \revrep{.}{with (a) one and (b) two \glspl{dsa} interfering.}
        A budget imbalance of 1:1 \dr{and a period longer than the application's runtime is selected.}
        We vary the fragmentation granularity.
        \dr{The applications contain an initialization, \emph{init}, and a \emph{main} phase.
        \emph{iso} denotes the isolated, \emph{unr} the unregulated performance.}
    }
    \label{fig:taclebench}
\end{figure}

To evaluate the functional performance of {\erealm} connected in front of the \gls{eth} peripheral in {\carfield}, we inject \gls{axi4} transaction faults within the \gls{eth} peripheral by either delaying to accept requests or stalling responses.
This behavior may occur due to a full transmission buffer or the \gls{eth} device failing to send the requested data.

We define the \gls{wcdt} as the time between the occurrence of a fault and the {\erealm} unit detecting it.
Most protocol errors, e.g., a wrong \gls{tid} or a superfluous handshake, are detected instantaneously.
To derive the \gls{wcdt}, we thus consider a timeout event.
As explained in \Cref{sec:arch:erealm}, an \gls{axi4} transaction is tracked in different stages, each having a dedicated time budget assigned.
The \gls{wcdt} is equal to the time of the largest budget configured for all stages.
For most transactions, the longest stage monitors the read or write data beats (\emph{r/w\_first to r/w\_last}), see \Cref{fig:erealm-stage}.
A timeout happens when a peripheral sends the first read or write \emph{r/w\_first} but never continues to issue any more beats; the {\erealm} unit detects this after the stage's budget is depleted.

In {\carfield}, we set the budgets for all stages, but the \emph{r/w\_first to r/w\_last} to \SI{20}{cycles}.
As the \gls{eth} \gls{ip} supports bursts up to \SI{256}{beats} in length, we set the budget for the read and write monitoring to \SI{300}{cycles}, corresponding to the \gls{wcdt} of the \gls{eth} \gls{ip}.
{\carfield} implements fast virtualized interrupt support through a core-local interrupt controller, enabling best-in-class interrupt responses of~\SI{100}{cycles} on the {CVA6} host core, ensuring quick and agile system reactions to disturbances.
The worst-case latency, from the fault occurring to the core reacting, is thus at most \SI{400}{cycles}.
The {\erealm} unit can be configured to automatically reset the faulty subordinate device within two cycles from fault detection, preparing it to resume operation immediately once the core is informed.

Thanks to the stage-level tracking of {\erealm} and the fast interrupt support of {\carfield}, we can inform the core and reset the subordinate in as low as \SI{100}{cycles} with a \gls{wcdt} of \SI{400}{cycles} for the \gls{eth} \gls{ip} in {\carfield}.
This response time is substantially lower than waiting for a deadline miss or a system-wide watchdog reset due to an interconnect stall from a faulty subordinate.

\subsection{Case Study on {\carfield}}
\label{sec:case:bmk-eval}

We evaluate the performance of {\axirealm} using heterogeneous benchmark applications on {\carfield}.
We combine a \gls{ml} inference application running on \rev{either one or both of} {\carfield}'s \lbb{machine-learning}-optimized \revrep{DSA}{\glspl{dsa}} while executing time-critical applications from \emph{TACLeBench}~\cite{TACLeBench} on the platform's host {RISC-V} cores.
To simulate a truly heterogeneous application including communication and coarse-grained synchronization, both the \gls{dsa} and the time-critical tasks access {\carfield}'s shared \gls{l2} \gls{spm} memory.
{TACLeBench} provides two real-world applications: \emph{lift} and \emph{powerwindow}; the first mimics the controller of an industrial elevator, the latter one of the four car windows controlled by the driver and the passenger~\cite{TACLeBench}.

We use {\axirealm} to control and monitor the data streams injected by the \gls{dsa}\rev{s} and the {RISC-V} cores to restore the performance of the {TACLeBench} applications.
We keep the budget between the\revdel{ two} actors equal and large enough to restrict neither the accelerator's \gls{dma}\rev{s} nor the cores from accessing \gls{l2}.
We chose a reasonably small period to mitigate any imbalance issues presented in \Cref{sec:case:per-bud}.
The fragmentation is swept between \emph{one}, fairest, and \emph{255} (no fragmentation).

For compute-bound applications (\emph{lift}), we observe no interference of the \gls{dsa}\rev{s} with the execution of the critical program.
{\axirealm} does not introduce any measurable overheads in these cases.
{\axirealm} can be rapidly disabled, see \Cref{sec:arch:prog}, should a non-memory-intensive task be executed on the platform, fully eliminating any dynamic power overhead of the {\axirealm} units.
The \emph{powerwindow} task faces up to \revrep{7~x}{\SI{9.7}{\times}} interference from the DSAs in shared memory. (\Cref{fig:taclebench}).
{\axirealm} reduces this delay by achieving \rev{in both cases} up to \SI{99}{\percent} of isolated performance with full fragmentation.

\section{Related Work}
\label{sec:relwrk}

We structure {\axirealm}'s related work into two parts: \Cref{sec:relwrk:irealm} compares the {\irealm} unit to real-time interconnects and regulation modules and \Cref{sec:relwrk:erealm} compares the {\erealm} unit to \gls{sota} transaction monitoring units.

\subsection{Real-time Extensions: \irealm}
\label{sec:relwrk:irealm}

The \gls{sota} of real-time interconnect extensions can be divided into two fundamental design strategies: \emph{drop-in regulation modules}, which are integrated between the managers and the interconnect itself, or intrusive \emph{interconnect architecture customizations}.
The latter strategy profoundly changes the interconnect's internal structure, intertwining the enhancements with a given memory system architecture and, thus, with a given system~\lbb{\cite{RETIS_AXI_HYPER_CONNECT, AXIICRT_ARM, BSC_ARM_QOS400}}.

\subsubsection{Drop-in Regulation Modules}
\label{subsec:relwork_helpers}

Credit-based mechanisms are commonly introduced at the boundary of existing interconnect configurations to impose spatial and temporal bounds on non-coherent, on-chip interconnect networks.

Pagani~{\etal} and Restuccia~{\etal} analyze and address the problem of multiple \glspl{dsa} either competing for bandwidth or causing interference in heterogeneous, \gls{axi4}-based \gls{fpga} \glspl{soc}.
They propose three units to mitigate contention.
The \emph{\gls{axi4} budgeting unit} (ABU)~\cite{RETIS_BANDWIDTH_RESERVATION} extends the concept of inter-core memory reservation established in \glspl{mpsoc} to heterogeneous \glspl{soc}.
The {ABU} uses counter-based budgets and periods assigned for each manager in the system, reserving a given bandwidth to each manager.
The \emph{\gls{axi4} burst equalizer} (ABE)~\cite{RETIS_IS_BUS_ARBITER_FAIR} tackles unfair arbitration by limiting the nominal burst size and a maximum number of outstanding transactions for each manager.
The \emph{Cut and Forward} (C\&F)~\cite{RETIS_CUT_FORWARD} unit prevents ahead-of-time bandwidth reservations by holding back transactions until they can certainly be issued.  
Our {\axirealm} architecture tackles these challenges while optimizing the design to be suitable for high-performance systems and use cases.
{\axirealm} adds only one cycle of latency (\Cref{sec:arch}) and extremely low area overhead (\Cref{sec:ip-results}).

Farshchi et al.~\cite{BRU} propose the \emph{Bandwidth Regulation Unit} (BRU), a \gls{hw} module aimed at reducing the regulation overheads of \gls{sw} approaches.
Designed for coherent multi-core \glspl{soc}, BRU manages memory traffic per core.
Akin to {\axirealm}, it employs a time-slicing approach, albeit with one global period shared by all domains.
The design can only regulate the maximum bandwidth, whose size is fixed to the dimension of a cache line, while the number of memory access transactions is user-configurable.
Implemented in a \SI{7}{\nano\meter} node, BRU adds minimal logic overhead ($<$0.3\%) and reduces the maximum achievable frequency by $<$2\%.
Furthermore, BRU can independently control the write-back traffic to the main memory, mitigating write-read imbalances.
Unlike \gls{sw} techniques~\cite{Bechtel2019DenialofServiceAO}, this functionality requires significant \gls{hw} modifications to the cache hierarchy.
We argue that it would favor a cache-centric partitioning strategy over an auxiliary module for the system bus.

A key aspect of achieving temporal isolation for shared resources involves extracting significant data from functional units during \gls{vv}.
This step is essential for determining an optimal upper limit for resource usage during operation.
Cabo~{\etal}~\cite{9774515, andreu2025expandingsafesu} propose \emph{SafeSU}, a minimally invasive statistics unit.
SafeSU tracks inter-core interference in \glspl{mpsoc} using dedicated counters.
Instead of limiting the number of transferred bytes, the maximum-contention control unit (MCCU) allocates timing interference quotas to each manager core in clock cycles.
Whenever the allocated quota is exceeded, an interrupt is raised.
SafeSU uses temporal information, including contention, request duration, and interference quota, as knobs to enhance traffic observability and enforce controllability.
Furthermore, the mechanism addresses interference exclusively in symmetric, general-purpose, multi-core systems.
{\axirealm} leverages spatial and temporal information for traffic regulation (i.e., bandwidth reservation and time slicing) and extends the monitoring capabilities to heterogeneous \glspl{mcs} comprising of real-time-critical, general-purpose, and high-performance domain-specific managers.

\subsubsection{Interconnect Customization}
\label{subsec:relwork_bus_redesign}

Restuccia~{\etal}~\cite{RETIS_AXI_HYPER_CONNECT} propose \emph{HyperConnect}, a custom \gls{axi4}-based functional unit block for virtualized \gls{fpga}-\glspl{soc}.
While being the closest \gls{sota} to {\axirealm}, HyperConnect does not tackle ahead-of-time bandwidth reservation issues caused by a slow manager stalling the interconnect (\Cref{sec:arch:buf}).

Recently, Jiang~{\etal} introduced \emph{AXI-IC$^{RT}$}~\cite{AXIICRT_ARM}, one of the first end-to-end \gls{axi4} microarchitectures tailored for real-time use cases.
AXI-IC$^{RT}$ leverages the \gls{axi4} user signal to assign priorities and introduces a dual-layer scheduling algorithm for the dynamic allocation of budget and period to each manager during runtime. 
To prevent request starvation on low-priority managers, {\axirealm} does not depend on the concept of priority, but rather on a credit-based mechanism and a \emph{burst splitter} to distribute the bandwidth according to the real-time guarantee of the \gls{soc}. 
While AXI-IC$^{RT}$ supports several budget reservation strategies, it limits the assessment to managers with equal credit (bandwidth).
Finally, from an implementation angle, the design strategy followed by AXI-IC$^{RT}$ adds extensive buffering to the microarchitecture to create an observation window for early service of incoming transactions based on priorities.
Overall, HyperConnect and AXI-IC$^{RT}$ lack monitoring capabilities to track traffic statistics.

In industry, Arm's \emph{CoreLink QoS-400} is widely integrated into modern \gls{fpga}-\glspl{soc} to manage contention using the QoS signal defined in the \gls{axi4} and AXI5 specifications.
However, QoS-400 has several limitations, as analyzed in~\cite{BSC_ARM_QOS400}.
One significant drawback is its intrusiveness; for instance, in a {Zynq Ultrascale+} \gls{fpga}, the authors report the need to coordinate over 30 {QoS} points to effectively control traffic~\cite{BSC_ARM_QOS400}.

\begin{table*}[t]
    \centering
        \centering
        \caption{%
            \Gls{sota} comparison of {\axirealm}.%
        }%
        \label{tab:soa}
        \centering
        \resizebox{1\linewidth}{!}{%
            \begin{threeparttable}
                \renewcommand{\arraystretch}{1.2}
                \begin{tabular}{@{}lllccccccccccc@{}}
                    \toprule
                    &
                    &
                    &
                    \tlb{Budget/}{Time}{Slicing} &
                    \textbf{Granularity} &
                    \dlb{Fair}{Arbitration} &
                    \tlb{Bandwidth}{Reservation}{Prevention} &
                    \dlb{Manager}{Isolation} &
                    \textbf{Statistics} &
                    \dlb{Protocol}{Checking} &
                    \dlb{Fault}{Handling} &
                    \dlb{Multiple}{Outstanding} &
                    \dlb{Target}{Technology} &
                    \dlb{Area}{Overhead} \\
                    \arrayrulecolor{ieee-dark-black-100}\midrule
                    \multirow{9}{*}{\vspace{-2.5cm}\rot{\textbf{\irealm}}} &
                    \multirow{5}{*}{\vspace{-2cm}\rot{\dlb{Regulation}{Helpers}}} &
                    ABU~\cite{RETIS_BANDWIDTH_RESERVATION} &
                    \dl{Period-}{based~\tnote{a}} &
                    \tl{One}{subordinate}{only} &
                    \xmark &
                    \xmark &
                    \xmark &
                    \xmark &
                    \na &
                    \tl{Ideal}{subordinate}{assumed} &
                    \xmark &
                    \dl{Xilinx}{FPGA} &
                    \dl{\SI{715}{LUT}}{\SI{908}{FF}}~\tnote{c} \\
                    &
                    &
                    ABE~\cite{RETIS_IS_BUS_ARBITER_FAIR} &
                    \xmark &
                    \na &
                    \dl{Transfer}{fragmentation} &
                    \xmark &
                    \xmark &
                    \xmark &
                    \na &
                    \tl{Ideal}{subordinate}{assumed} &
                    \cmark &
                    \dl{Xilinx}{FPGA} &
                    \dl{\SI{1130}{LUT}}{\SI{582}{FF}}~\tnote{c} \\
                    &
                    &
                    C\&F~\cite{RETIS_CUT_FORWARD} &
                    \xmark &
                    \na &
                    \xmark &
                    \dl{Write}{buffering} &
                    \xmark &
                    \xmark &
                    \na &
                    \tl{Ideal}{subordinate}{assumed} &
                    \na &
                    \dl{Xilinx}{FPGA} &
                    \dl{\SI{3088}{LUT}}{\SI{1467}{FF}}~\tnote{d,e} \\
                    &
                    &
                    SafeSU~\cite{9774515,andreu2025expandingsafesu} &
                    SW &
                    SW &
                    \na &
                    \xmark &
                    \xmark &
                    \tl{Bandwidth,}{latency,}{interference} &
                    \na &
                    \xmark &
                    \xmark &
                    \dl{Tech.-}{independent} &
                    \dl{\SI{83.1}{\kGE}~\tnote{f, g}}{\SI{223}{\kGE}~\tnote{f, h}} \\ %
                    &
                    &
                    BRU~\cite{BRU} &
                    \tl{Global}{period-}{based~\tnote{a}} &
                    \dl{Shared}{memory} &
                    \xmark &
                    \xmark &
                    \xmark &
                    \xmark &
                    \na &
                    \tl{Ideal}{subordinate}{assumed} &
                    \xmark &
                    \dl{Tech.-}{independent} &
                    \SI{57.2}{\kGE}~\tnote{b} \\ %
                    \arrayrulecolor{ieee-dark-black-40}\cdashline{2-14}
                    &
                    \multirow{4}{*}{\rot{\dlb{Interconn.}{Cust.}}} &
                    Hyperconnect~\cite{RETIS_AXI_HYPER_CONNECT} &
                    \dl{Period-}{based~\tnote{a}} &
                    \dl{Per}{subordinate} &
                    \dl{Transfer}{fragmentation} &
                    \dl{Write}{buffering} &
                    \xmark &
                    \xmark &
                    \na &
                    \tl{Ideal}{subordinate}{assumed} &
                    \cmark &
                    \dl{Xilinx}{FPGA} &
                    \dl{\SI{3020}{LUT}}{\SI{1289}{FF}}~\tnote{c} \\
                    &
                    &
                    AXI-IC$^{RT}$~\cite{AXIICRT_ARM} &
                    \dl{Period-}{based~\tnote{a}} &
                    \dl{Per}{subordinate} &
                    \dl{Transfer}{buffering} &
                    Buffering &
                    \xmark &
                    \xmark &
                    \na &
                    \xmark &
                    \cmark &
                    FPGA &
                    \dl{\SI{4745}{LUT}}{\SI{4184}{FF}}~\tnote{i} \\
                    &
                    &
                    QOS-400~\cite{BSC_ARM_QOS400} &
                    \dl{Prio.-}{based~\tnote{a}} &
                    \dl{Per}{subordinate} &
                    \xmark &
                    \xmark &
                    \xmark &
                    \xmark &
                    \na &
                    \na &
                    \cmark &
                    \dl{Tech.-}{independent} &
                    \na \\
                    \arrayrulecolor{ieee-dark-black-40}\midrule
                    \multirow{6}{*}{\vspace{-1.6cm}\rot{\textbf{\erealm}}} &
                    \multirow{6}{*}{\vspace{-1.5cm}\rot{\dlb{Subordinate}{Guarding}}} &
                    SP805~\cite{sp805_arm} &
                    \na &
                    \dl{Entire}{system} &
                    \na &
                    \na &
                    \xmark &
                    \todo{\cmark} &
                    \xmark &
                    \tl{IRQ,}{Global}{reset}&
                    \xmark &
                    \dl{Tech.-}{independent} &
                    \na \\
                    &
                    &
                    Synopsys~\cite{synopsys_smart} &
                    \na &
                    \dl{Per}{subordinate} &
                    \na &
                    \na &
                    \xmark &
                    \todo{\cmark} &
                    \xmark &
                    \na &
                    \xmark &
                    \dl{Tech.-}{independent} &
                    \na \\
                    &
                    &
                    AMD~\cite{amd_axi_perf} &
                    \na &
                    \dl{Per}{subordinate} &
                    \na &
                    \na &
                    \xmark &
                    \todo{\cmark} &
                    \xmark &
                    \na &
                    \xmark &
                    \dl{Tech.-}{independent} &
                    \na \\
                    &
                    &
                    Ravi~\etal~\cite{ravi2014design} &
                    \na &
                    \dl{Per}{subordinate} &
                    \na &
                    \na &
                    \xmark &
                    \todo{\cmark} &
                    \xmark &
                    \na &
                    \xmark &
                    \dl{Tech.-}{independent} &
                    \SI{8.86}{\kGE}~\tnote{j} \\ %
                    &
                    &
                    Kyung~\etal~\cite{kyung2007performance} &
                    \na &
                    \dl{Per}{subordinate} &
                    \na &
                    \na &
                    \xmark &
                    \todo{\cmark} &
                    \xmark &
                    \na &
                    \xmark &
                    \dl{FPGA}{platform} &
                    \na \\
                    &
                    &
                    Lee~\etal~\cite{lee2014reconfigurable} &
                    \na &
                    \dl{Per}{subordinate} &
                    \na &
                    \na &
                    \xmark &
                    \todo{\cmark} &
                    \todo{\cmark} &
                    \dl{Logging-}{only} &
                    \xmark &
                    \dl{FPGA}{platform} &
                    \na \\
                    &
                    &
                    AXIChecker~\cite{chen2010synthesizable} &
                    \na &
                    \dl{Per}{subordinate} &
                    \na &
                    \na &
                    \na &
                    \xmark &
                    Performance &
                    \xmark &
                    \xmark &
                    \dl{Tech.-}{independent} &
                    \SI{70.7}{\kGE} \\
                    \arrayrulecolor{ieee-dark-black-40}\midrule
                    \multicolumn{2}{c}{\textbf{\irealm}} &
                    \multirow{2}{*}{\emph{\axirealm~[Ours]}} &
                    \dl{Period-}{based~\tnote{a}} &
                    \tlb{Configurable}{subordinate}{regions} &
                    \dl{Transfer}{fragmentation} &
                    \dl{Write}{buffering} &
                    \dlb{Per}{manager} &
                    \tlb{Per-region}{bandwidth,}{latency} &
                    \na &
                    \multirow{2}{*}{\tlb{IRQ,}{per-subordinate}{reset}} &
                    \dlb{Throttling}{mechanism} &
                    \multirow{2}{*}{\dl{Tech.-}{independent}} &
                    \dlb{As low as}{\SI{5}{\kGE}} \\
                    \multicolumn{2}{c}{\textbf{\erealm}} &
                    &
                    \na &
                    \dl{Per}{subordinate} &
                    \na &
                    \na &
                    \dlb{Per}{subordinate} &
                    \dlb{Latency}{fine-granular} &
                    \todo{\cmark} &
                    &
                    \cmark &
                    &
                    \dlb{As low as}{\SI{15}{\kGE}} \\
                    \arrayrulecolor{ieee-dark-black-100}\bottomrule
                \end{tabular}

                \begin{tablenotes}[para, flushleft]
                    \item[a] in hardware
                    \item[b] assuming \SI{0.08748}{\micro\metre\squared} for \SI{1}{\GE} (\emph{NAND2x1\_ASAP7\_75t\_R})
                    \item[c] Xilinx Zynq-7020
                    \item[d] Xilinx ZCU102
                    \item[e] C=4
                    \item[f] assuming \SI{1.28}{\micro\metre\squared} for \SI{1}{\GE} 
                    \item[g] SafeSU
                    \item[h] SafeSU-2
                    \item[i] Xilinx VC709
                    \item[j] assuming \SI{0.718}{\micro\metre\squared} for \SI{1}{\GE} 
                \end{tablenotes}
            \end{threeparttable}
        }
\end{table*}

\subsection{Subordinate Guarding: \erealm}
\label{sec:relwrk:erealm}

All the works described in \Cref{sec:relwrk:irealm} assume \emph{perfect} subordinate behavior, providing a response within a bounded time, and focus on the manager side without considering malfunctioning or misbehaving subordinates.
Our {\erealm} unit tackles these challenges by monitoring and guarding subordinate devices.
Transaction monitoring and guarding are crucial in studying security, performance analysis, fault detection, and system reliability.
With the {\erealm} unit, we use these established concepts in real-time memory interconnect systems.

Arm's \emph{SP805 Watchdog}~\cite{sp805_arm} is primarily designed for fault detection and system protection by safeguarding the \glspl{soc} against \gls{sw} malfunctions due to unresponsive or runaway processes.
The operating system has to reset an internal counter regularly; if it becomes unresponsive, SP805 can either emit an interrupt or reset the entire system.
In contrast, {\erealm} provides a \gls{hw} solution that monitors every subordinate access, reducing fault detection latency.
We allow dynamic time budgeting of each subordinate device's transaction phases and thus support tight latency bounds for each device individually.
Unlike SP805, our approach allows the selective reset of the non-responsive subordinate device within a single cycle, leaving the rest of the system operational.

We identify multiple units specialized in monitoring subordinate devices.
With Synopsys' \emph{Smart Monitor}~\cite{synopsys_smart} and AMD's \emph{\gls{axi4} Performance Monitor}~\cite{amd_axi_perf}, industry provides performance monitoring solutions for \gls{axi4} buses and subordinate devices.
These units monitor bus traffic and compute key performance metrics, such as data byte count, throughput, and latency.
In academia, Ravi~{\etal} present a \emph{Bus Monitor}~\cite{ravi2014design} and Kyung~{\etal} describe their \emph{Performance Monitoring Unit} (PMU)~\cite{kyung2007performance} to capture key performance metrics such as transaction count, transfer size, and latency distributions for \gls{axi4} transactions through \gls{hw} counters.
Compared to {\erealm}, neither support multiple outstanding transactions nor provide detailed, \emph{stage-specific} transaction insights.
This limits their use in heterogeneous \glspl{soc}, where high-performance \glspl{dsa} emit complex transactions, and detailed performance reports of individual transactions are required.
Delayed or missing responses are not the only critical fault a subordinate device can experience.
Lee~{\etal}~\cite{lee2014reconfigurable} describe a \emph{Reconfigurable Bus Monitor Tool Suite} for on-chip monitoring of \glspl{soc}.
The suite offers a \emph{Bus Monitor IP} designed to monitor the device's performance and check key protocol properties.
For the latter, it verifies simple specification-compliance, but unlike {\erealm}, it does not offer any protection in multi-{ID} scenarios with multiple outstanding transactions, e.g., \gls{tid} mismatch.
Chen~{\etal} developed \emph{AXIChecker}~\cite{chen2010synthesizable}, a rule-based, synthesizable protocol checker enforcing 44 rules ensuring managers and subordinates operate protocol-compliant.
Compared to {\erealm}, it can log protocol issues but lacks performance monitoring and reaction capabilities.

\subsection{Final Remarks: \axirealm}
\label{sec:relwrk:axirealm}

\lbb{A distinctive aspect of} \emph{{\axirealm}} \lbb{is} in its modular design.
It seamlessly combines ingress monitoring and throttling to ensure real-time behavior across managers with egress monitoring and guarding to guarantee timely responses from subordinate devices.
Its transparent and modular design requires minimal changes to the system, and its compatibility with many well-tested, silicon-proven crossbars and interconnects eases integration and verification.  
Most \gls{sota} solutions explicitly restrict the design and evaluation on \gls{fpga} platforms, lacking support for \glspl{asic}.
Our technology-independent approach provides in-system and \gls{ip}-level gate-level characterization in a modern technology node, facilitating \gls{sota} comparisons.

\section{Conclusion}
\label{sec:conclusion}

We present {\axirealm} a lightweight, minimally invasive, architecture-independent, open-source interconnect extension to enable real-time behavior in high-performance interconnects used in heterogeneous systems.

The {\irealm} unit offers an effective solution for monitoring and moderating manager traffic during interference scenarios on a shared interconnect.
\fr{It provisions isolation and enforces real-time guarantees to managers executing critical tasks in heterogeneous systems.}
Integrated into {\carfield}, an open-source \gls{mcs} research platform, we achieve \todo{\SI{68}{\percent}} of the ideal performance in memory-bound applications, \lbb{massively reducing the memory access latency by \SI{24}{\times},} while incurring less than \SI{2}{\percent} of additional area.
When distributing the budget in favor of the core, we achieve over \todo{\SI{95}{\percent}} of the isolated performance. 
Running applications from {TACLeBench}, we achieve over \SI{98}{\percent} of the isolated performance.

With the {\erealm} unit, we include an effective \gls{hw}-based solution to gracefully handle malfunctioning subordinates individually without stalling or locking the rest of the interconnect or the system.
The unit monitors the transaction latency and protocol correctness of each guarded subordinate, being able to inform the application-class core in as low as \SI{100}{cycles}, handshake open transaction, and reset the device should a transaction be overly delayed or the subordinate malfunctioning ensuring timely responses and real-time guarantees.

\section*{Acknowledgments}
This work has received funding from the Swiss State Secretariat for Education, Research, and Innovation (SERI) under the SwissChips initiative.
%

%
\input{main.bbl}

%
\newcommand{\missingbio}{has not yet added their bio. This is just a placeholder.}
\newcommand{\lucaphd}[1]{#1 is currently pursuing a Ph.D. degree in the Digital Circuits and Systems group of Prof.\ Benini.}
\newcommand{\ethgrad}[3]{received #1 B.Sc. and M.Sc. degrees in electrical engineering and information technology from ETH Zurich in #2 and #3, respectively.}
\newcommand{\lucagrad}[2]{completed #1 Ph.D. in the Digital Circuits and Systems group of Prof.\ Benini in #2.}
\newcommand{\researchinterests}[1]{research interests include #1.}

\vspace{0.0cm}
\begin{IEEEbiography}[%
    {\includegraphics[width=1in,height=1.25in,clip,keepaspectratio]{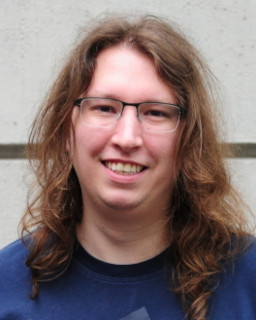}}%
    ]{Thomas Benz}
    (Graduate Student Member, IEEE)
    \ethgrad{his}{2018}{2020}
    \lucaphd{He}
    His
    \researchinterests{energy-efficient high-performance computer architectures, memory interconnects, data movement, and the design of \acrshortpl{asic}}
\end{IEEEbiography}\vspace{-0.33cm}

\begin{IEEEbiography}[%
    {\includegraphics[width=1in,height=1.25in,clip,keepaspectratio]{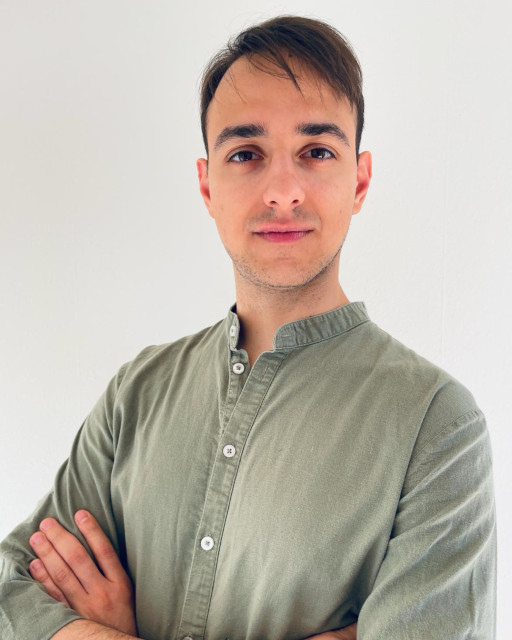}}%
    ]{Alessandro Ottaviano}
    (Graduate Student Member, IEEE)
    received the B.Sc. in Physical Engineering from Politecnico di Torino, Italy, and the M.Sc. in Electrical Engineering as a joint degree between Politecnico di Torino, Grenoble INP-Phelma and EPFL Lausanne, in 2018 and 2020 respectively. %
    \lucaphd{He}
    His
    \researchinterests{real-time and predictable computing systems and energy-efficient processor architecture}
\end{IEEEbiography}\vspace{-0.33cm}

\begin{IEEEbiography}[%
    {\includegraphics[width=1in,height=1.25in,clip,keepaspectratio]{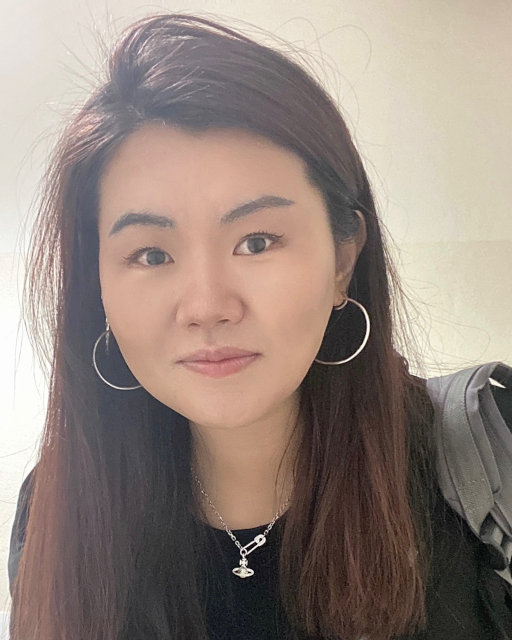}}%
    ]{Chaoqun Liang}
    (Graduate Student Member, IEEE)
    received the M.Sc degree in electronic engineering from University of Bologna, Italy and Technical University of Munich, Germany in 2023. She is currently pursuing a Ph.D degree with the Department of Electrical, Electronic and Information Engineering, University of Bologna, Italy. Her main research interests are on-chip and off-chip communication.
\end{IEEEbiography}\vspace{-0.33cm}

\vfill
\newpage

\begin{IEEEbiography}[%
    {\includegraphics[width=1in,height=1.25in,clip,keepaspectratio]{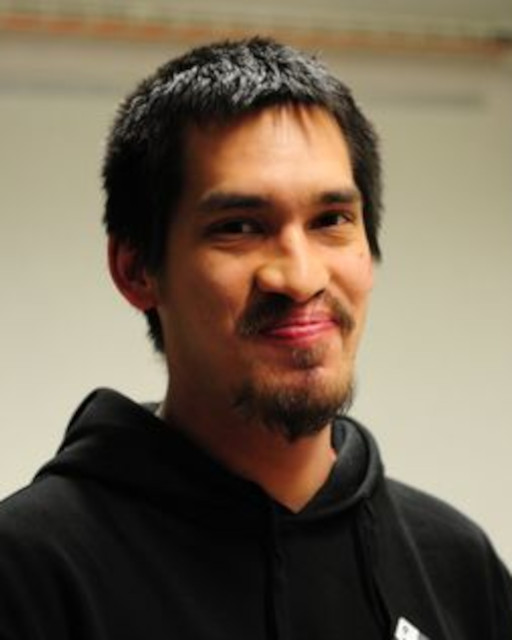}}%
    ]{Robert Balas}
    (Graduate Student Member, IEEE)
    \ethgrad{his}{2015}{2017}
    \lucaphd{He}
    His
    \researchinterests{real-time computing, compilers, and operating-systems}
\end{IEEEbiography}\vspace{-0.33cm}

\begin{IEEEbiography}[%
    {\includegraphics[width=1in,height=1.25in,clip,keepaspectratio]{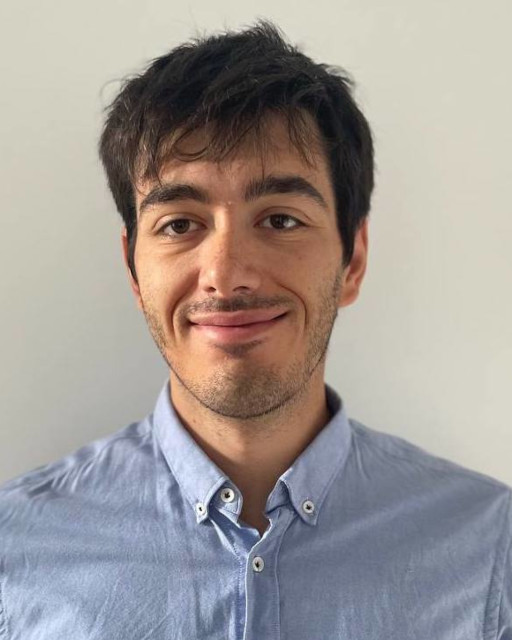}}%
    ]{Angelo Garofalo}
    (Member, IEEE) received his Ph.D. degree from the University of Bologna, Italy, in 2022. He is currently a Junior Assistant Professor (RTD-A) at the University of Bologna, Italy, and a postdoctoral researcher at ETH Zurich, Switzerland. His research focuses on reliable and time-predictable computing systems, heterogeneous edge AI architectures, and energy-efficient multi-core processors. He has published more than 30 peer-review papers and he is recipient of a best paper award at IEEE ISVLSI 2023.
\end{IEEEbiography}\vspace{-0.33cm}

\begin{IEEEbiography}[%
    {\includegraphics[width=1in,height=1.25in,clip,keepaspectratio]{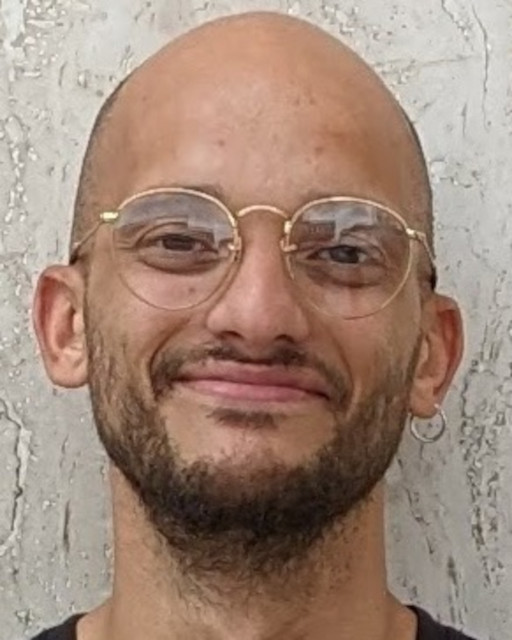}}%
    ]{Francesco Restuccia}
    (Member, IEEE) received his Ph.D. degree from Scuola Superiore Sant'Anna Pisa, Italy, in 2021. Since 2022, he is a postdoctoral researcher at the University of California San Diego. His research interests include hardware security, safety, and timing predictability on heterogeneous platforms. 
\end{IEEEbiography}\vspace{-0.33cm}

\begin{IEEEbiography}[%
    {\includegraphics[width=1in,height=1.25in,clip,keepaspectratio]{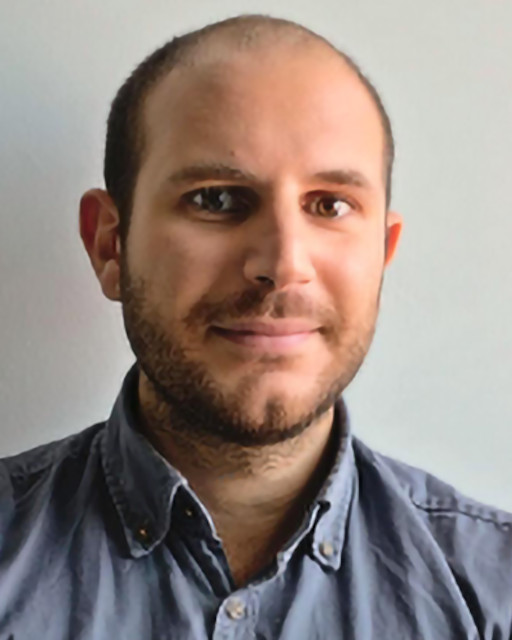}}%
    ]{Alessandro Biondi}
    (Member, IEEE) received his graduate (cum laude) degree in computer engineering from the University of Pisa, Italy, and his Ph.D. degree in computer engineering from Scuola Superiore Sant’Anna. He is an Associate Professor with the Real-Time Systems Laboratory, Scuola Superiore Sant’Anna. His research interests include designing and implementing real-time operating systems and hypervisors, schedulability analysis, cyberphysical systems, synchronization protocols, and component-based design for real-time multiprocessor systems.
\end{IEEEbiography}\vspace{-0.33cm}

\begin{IEEEbiography}[%
    {\includegraphics[width=1in,height=1.25in,clip,keepaspectratio]{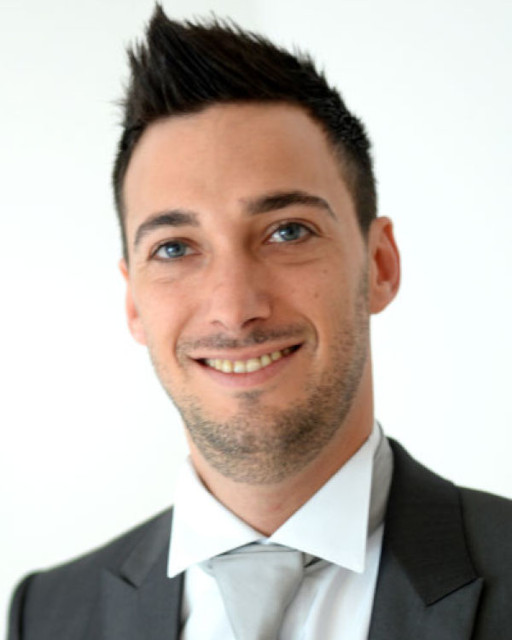}}%
    ]{Davide Rossi}
    (Senior Member, IEEE) received the Ph.D. degree from the University of Bologna, Bologna, Italy, in 2012. He has been a Post-Doctoral Researcher with the Department of Electrical, Electronic and Information Engineering “Guglielmo Marconi,” University of Bologna, since 2015, where he is currently an Associate Professor. His research interests include energy-efficient digital architectures.
\end{IEEEbiography}\vspace{-0.33cm}

\begin{IEEEbiography}[%
    {\includegraphics[width=1in,height=1.25in,clip,keepaspectratio]{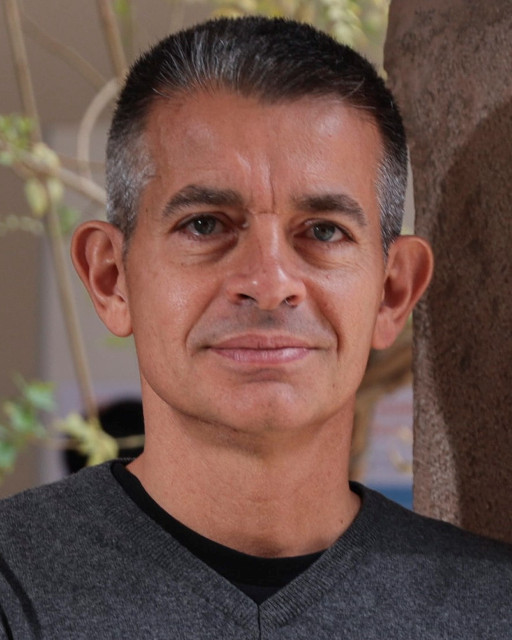}}%
    ]{Luca Benini}
    (Fellow, IEEE) 
    holds the chair of digital Circuits and systems at ETHZ and is Full Professor at the Università di Bologna.
    He received a PhD from Stanford University.
    His research interests are energy-efficient parallel computing systems, smart sensing micro-systems, and machine learning hardware.
    He is a Fellow of the IEEE, of the ACM, a member of the Academia Europaea, and of the Italian Academy of Engineering and Technology.
\end{IEEEbiography}\vspace{-0.33cm}

\vfill

\end{document}